\documentclass[12pt]{article}
\usepackage{authblk}
\usepackage{geometry}
\usepackage{amsthm}
\usepackage{mathrsfs}
\usepackage{customcommands}
\usepackage{bm}
\usepackage{Chrisstyle}

\usepackage{amsmath}
\usepackage{amssymb}
\usepackage{braket}
\usepackage[normalem]{ulem}
\usepackage{color}
\usepackage{ulem}
\usepackage{mathtools}
\usepackage{tensor} 
\usepackage{overpic} 
\usepackage{breqn} 
\usepackage{subcaption}
\usepackage{float}
\usepackage[export]{adjustbox}

\DeclareMathOperator{\tr}{tr}

\graphicspath{{./fig/}{./fig/sec3/}}

\theoremstyle{definition}

\theoremstyle{remark}

\title{The Page Curve for Reflected Entropy}

\author[1]{Chris Akers,}
\author[2]{Thomas Faulkner,}
\author[2]{Simon Lin}
\author[3]{and Pratik Rath}

\affiliation[1]{Center for Theoretical Physics,\\
Massachusetts Institute of Technology, Cambridge, MA 02139, USA}
\affiliation[2]{Department of Physics, University of Illinois,\\ 1110 W. Green St., Urbana, IL 61801-3080, USA}
\affiliation[3]{Department of Physics, University of California, Santa Barbara, CA 93106, USA}

\emailAdd{cakers@mit.edu}
\emailAdd{tomf@illinois.edu}
\emailAdd{shanlin3@illinois.edu}
\emailAdd{rath@ucsb.edu}

\abstract{We study the reflected entropy $S_R$ in the West Coast Model, a toy model of black hole evaporation consisting of JT gravity coupled to end-of-the-world branes. We demonstrate the validity of the holographic duality relating it to the entanglement wedge cross section away from phase transitions. Further, we analyze the important non-perturbative effects that smooth out the discontinuity in the $S_R$ phase transition. By performing the gravitational path integral, we obtain the reflected entanglement spectrum analytically. The spectrum takes a simple form consisting of superselection sectors, which we interpret as a direct sum of geometries, a disconnected one and a connected one involving a closed universe. We find that area fluctuations of $O(\sqrt{G_N})$ spread out the $S_R$ phase transition in the canonical ensemble, analogous to the entanglement entropy phase transition. We also consider a Renyi generalization of the reflected entropy and show that the location of the phase transition varies as a function of the Renyi parameter.
}

\begin{document}
\maketitle

\section{Introduction}\label{sec:intro}

The black hole information problem has served as a beacon guiding us in the quest to understand quantum gravity \cite{Hawking:1974sw,Mathur:2009hf,Almheiri:2012rt,Almheiri:2013hfa}. 
Although a complete resolution still eludes us and might require a better understanding of the UV-complete theory of quantum gravity, significant progress has been made in recent years simply by taking the gravitational path integral seriously \cite{Penington:2019kki, Almheiri:2019qdq}. 

A commendable milestone in this endeavour is the calculation of the ``Page Curve'' using the semiclassical theory \cite{Penington:2019npb, Almheiri:2019psf}. In fact, in the so-called West Coast Model \cite{Penington:2019npb}, a toy model of black hole evaporation consisting of Jackiw-Teitelboim (JT) gravity coupled to end-of-the-world (ETW) branes, the detailed curve including effects near the phase transition were computed. 

\begin{figure}[h!]
    \centering \includegraphics[width=0.4\textwidth]{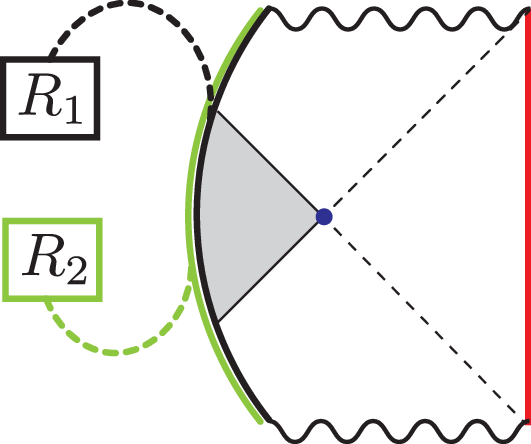}
    \caption{The Lorentzian description of the state we consider in the West Coast Model, a JT gravity black hole with an ETW brane. The ETW brane carries two sub-flavours, denoted black and green, that are entangled (dashed, coloured lines) with radiation systems $R_1$ and $R_2$ respectively. The extremal surface is denoted in purple and the island that dominates after the Page time is coloured gray.}
    \label{fig:west_coast_intro}
\end{figure}

Away from phase transitions, the Page curve can be computed using the quantum extremal surface (QES) formula \cite{Engelhardt:2014gca,Dong:2017xht,Akers:2021fut} which reads
\begin{equation}\label{eq:QES}
	S(A) = \min_{\gamma_A} \frac{\mathrm{Area}(\gamma_A)}{4 G_N} + S_{\text{bulk}}(\Sigma_A),
\end{equation}
where $S(A)$ is the entanglement entropy of subsystem $A$. $\Sigma_A$ is a partial Cauchy slice, whose domain of dependence is the entanglement wedge of $A$, such that $\partial \Sigma_A = A \cup \gamma_A$. In the West Coast Model, one considers a state with ETW branes carrying flavour indices that are entangled with an auxiliary radiation system $R=R_1\cup R_2$.\footnote{The division of $R$ into subsystems is a straightforward generalization, which will be useful for our computation of reflected entropy.} The model consists of two parameters: the horizon area $S_{BH}$ and the number of ETW brane flavour indices $k=k_1 k_2$. Tuning these parameters simulates black hole evaporation, and by applying \Eqref{eq:QES} to $R$, one finds the Page curve to be
\begin{equation}
	S(R) = \min(S_{BH},\log k),
\end{equation}
where the candidate quantum extremal surfaces are the bifurcation surface and the trivial surface respectively, as shown in \figref{fig:west_coast_intro}.

\begin{figure}
	\centering 
		\includegraphics[width=0.4
		\textwidth]{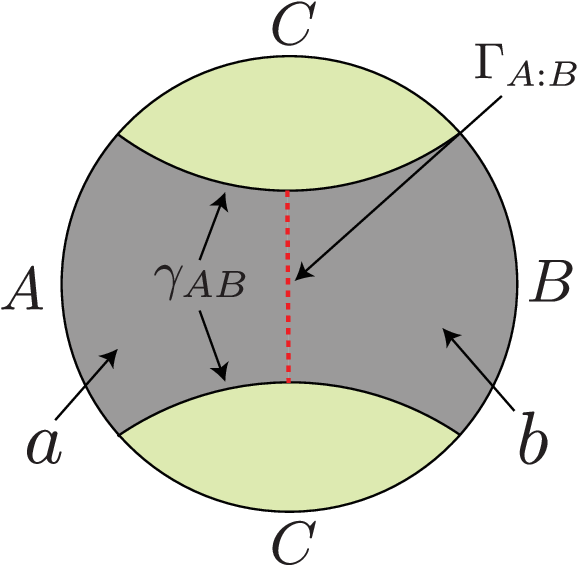}
	\caption{A spatial slice of AdS with $A$ and $B$ chosen to be two intervals. The figure depicts the entanglement wedge of $AB$ (gray), the entanglement wedge of $C$ (green), the RT surface $\gamma_{AB}$ and the entanglement wedge cross section $\Gamma_{A:B}$, which divides the entanglement wedge into regions $a$ and $b$ homologous to $A$ and $B$ respectively.} 
\label{fig:EW} \end{figure}

While the entanglement entropy is essentially the unique measure for bipartite entanglement in pure states, one would also like to understand the multipartite entanglement structure of the radiation. Various distinct measures of entanglement have been proposed for this purpose. 
Among these we shall focus on the reflected entropy $S_R(A:B)$, which seems to be particularly interesting for holography \cite{Dutta:2019gen, Akers:2019gcv}.

Given a density matrix $\rho_{AB}$, the reflected entropy $S_R(A:B)$ is defined by considering the canonically purified state $\ket{\sqrt{\rho}}_{AA^*BB^*}$ in a doubled Hilbert space where $A^*$ and $B^*$ are the mirror copies of $A$ and $B$. The reflected entropy is then defined as
\begin{equation}
	S_R(A:B) = S(AA^*)_{\ket{\sqrt{\rho}}}.
\end{equation}
The reflected entropy is proposed to be holographically computed by the entanglement wedge cross section \cite{Dutta:2019gen} and corresponding quantum corrections \cite{Chandrasekaran:2020qtn,Hayden:2021gno}, i.e.,
\begin{equation}\label{eq:EW}
	S_R(A:B) = \min_{\Gamma_{A:B}} \left[2 \frac{\mathrm{Area}(\Gamma_{A:B})}{4G_N} + S_{R,\text{bulk}}(a:b)\right],
\end{equation}
where $\Gamma_{A:B}$ is a surface dividing the entanglement wedge of $AB$ into portions $a$ and $b$, which are homologous to $A$ and $B$ respectively (see \figref{fig:EW}). The reflected entropy has already proven useful in understanding the requirement of tripartite entanglement in holographic states \cite{Akers:2019gcv, Hayden:2021gno}.

\begin{figure}
	\centering 
		\includegraphics[width=
		\textwidth]{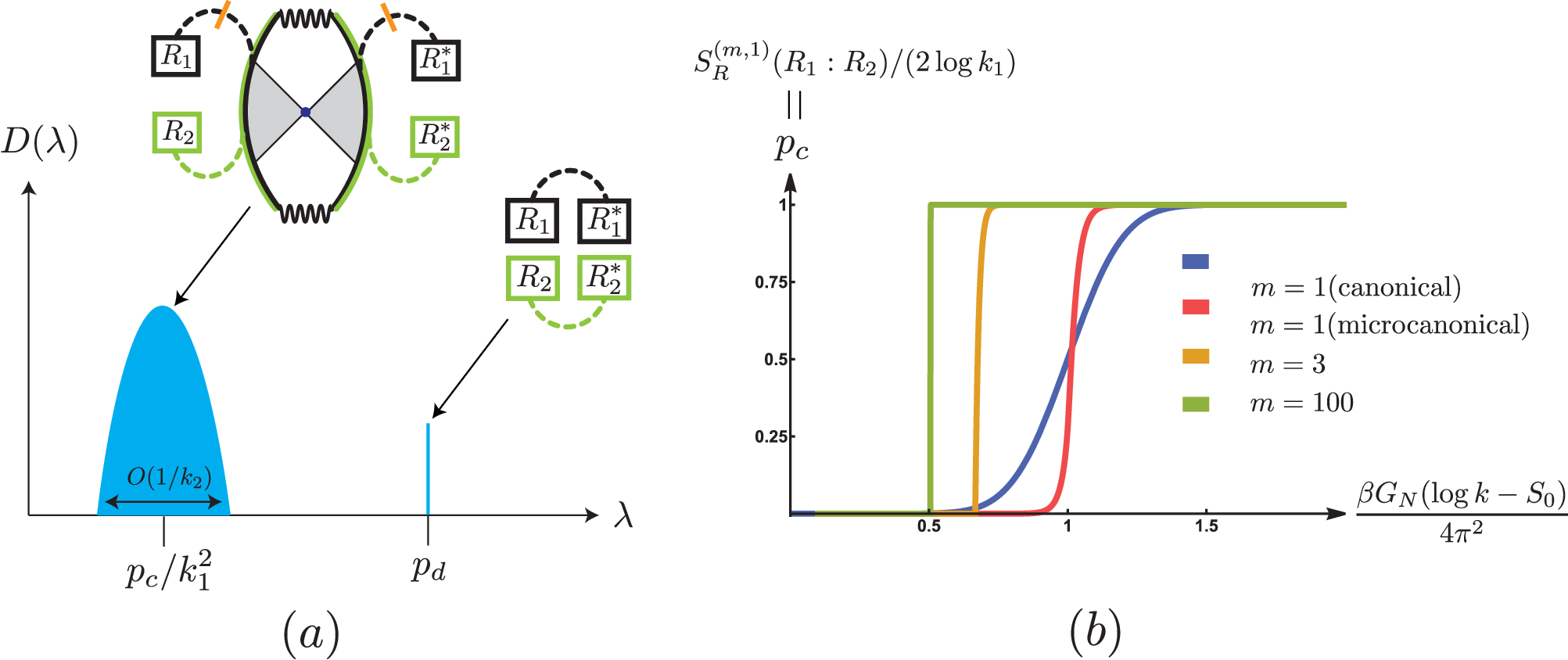} 
	\caption{The main result: (a) The reflected entanglement spectrum of $\rho_{AA^*}$ in the canonically purified state $\ket{\sqrt{\rho_{AB}}}$ is a mixture of two superselection sectors:  a single pole of weight $p_d$ corresponding to the disconnected purification, and a mound of approximately $k_1^2$ eigenvalues (assuming $k_1<k_2$) with weight $p_c$ corresponding to a connected purification involving a closed universe with the entanglement wedge cross section denoted in orange. (b) The probability of the connected purification $p_c$ as we vary $k$ across the Page transition. We show analytic plots for the microcanonical and canonical ensemble. The latter shows a spread in the phase transition of $O(\sqrt{G_N})$. We also show plots for $p_c$ when considering the $(m,1)$-Renyi reflected entropy; these undergo sharp transitions at $m$-dependent locations in the canonical ensemble.} 
\label{fig:results} \end{figure}

We analyze the reflected entropy between the radiation subsystems $R_1$ and $R_2$. Simply applying \Eqref{eq:EW} leads to
\begin{equation}\label{eq:wcsr}
	S_R(R_1:R_2) = \begin{cases}
		0 & k < \exp(S_{BH})\\
		2 \min(\log k_1,\log k_2) &  k>\exp(S_{BH})
	\end{cases}
\end{equation}
This follows from the two possible saddles obtained upon canonical purification: the disconnected geometry and the connected geometry depicted in \figref{fig:results}. 

In this paper we compute the reflected entropy precisely, demonstrating the validity of the holographic formula away from phase transitions. The fact that the gravitational path integral can be performed exactly allows us to compute the detailed behaviour of $S_R(R_1:R_2)$ including corrections to \Eqref{eq:wcsr} near the phase transition. We term this behavior the `Page curve for reflected entropy.' This is the main result of this paper, summarized by \figref{fig:results}.

In fact, we compute the entire reflected entanglement spectrum, which takes a very simple form: it consists of a mixture of two superselection sectors corresponding to the disconnected and connected geometries shown in \figref{fig:results}. The probabilities $p_d$ and $p_c$ of the two sectors are computable functions, that vary as we change $k$, leading to the phase transition in \Eqref{eq:wcsr}. Moreover, we find corrections to \Eqref{eq:wcsr} in a window of $\Delta \log k = O(\frac{1}{\sqrt{G_N}})$ near the phase transition that arise from fluctuations in the horizon area\footnote{In JT gravity, this corresponds to the value of the dilaton at the bifurcation surface.} (equivalently thermal fluctuations, in this model). This is analogous to a similar effect found in the entanglement entropy Page curve \cite{Vidmar:2017pak,Murthy:2019qvb,Penington:2019kki,Dong:2020iod,Marolf:2020vsi}, and we expect it to be a universal feature of holographic reflected entropy.

\subsection*{Overview}

In \secref{sec:setup}, we set up the stage for our analysis. In \secref{sub:wc}, we review the West Coast Model. In \secref{sub:sr}, we discuss basic aspects of the reflected entropy in this model. The proposed holographic answer arises from a leading saddle computation assuming replica symmetry. This calculation suffers from various issues, which motivates us to analyze the problem in more detail.

In \secref{sec:EE}, we review the computation of the entanglement spectrum in the West Coast Model. This serves as a warm-up for the resolvent trick which is used to obtain the spectrum in \secref{sub:resolvent_ee}. We then use the spectrum to compute the Renyi entropies in this model in \secref{sub:renyi}. Apart from being of interest on their own, the results for the Renyi entropies serve as an input for the reflected entropy calculation.

In \secref{sec:SR}, we analyze the reflected entanglement spectrum in the West Coast Model. We describe the resolvent trick in \secref{sub:resolvent_sr} to obtain a Schwinger-Dyson equation for the reflected entanglement spectrum. In \secref{sub:spectrum}, we solve this equation to obtain the spectrum and use it to analyze the reflected entropy. In \secref{sub:renyi_sr}, we go on to analyze a two-parameter Renyi generalization of the reflected entropy, called the $(m,n)$-Renyi reflected entropies.

In \secref{sec:disc}, we discuss these results and future directions.

\textit{Note:} This paper provides a complementary analysis to Ref.~\cite{Akers:2021pvd} where we analyze the reflected entropy phase transition in random tensor networks. 

\section{Preliminaries} 
\label{sec:setup}

In this section, we set up the background for our analysis. We will first review the West Coast Model in \secref{sub:wc}. In \secref{sub:sr}, we review aspects of the reflected entropy and discuss the proposed holographic dual.

\subsection{West Coast Model} 
\label{sub:wc}


The West Coast Model is a toy model that was used to derive the Page curve of entanglement entropy. We briefly describe the model here, for more details refer to Ref.~\cite{Penington:2019kki}. The model consists of JT gravity coupled to ETW branes. The action is given by
\begin{align}
	I &= I_{JT}+\mu\int_{\text{brane}}ds\\
	I_{JT}&=-\frac{S_0}{2 \pi}\left[\frac{1}{2} \int_{\mathcal{M}}\sqrt{g}R+\int_{\partial \mathcal{M}}\sqrt{h}K \right] - \left[\frac{1}{2} \int_{\mathcal{M}}\sqrt{g}\phi(R+2)+\int_{\partial \mathcal{M}}\sqrt{h}\phi K \right], 
\end{align}
where $S_0$ is the extremal entropy and $\mu$ is the mass of the brane. We will take both the above parameters to be large in our analysis. The ETW branes also possess a large number of flavours $k$, which we artificially divide into two sub-flavour indices that number $k_1$ and $k_2$ respectively. 

We consider a state where the ETW brane is entangled with two radiation bath systems $R_1$ and $R_2$ as
\begin{equation}\label{eq:state}
	\ket{\Psi} = \frac{1}{\sqrt{k}} \sum_{i=1}^{k_1}\sum_{j=1}^{k_2} \ket{i}_{R_1}\ket{j}_{R_2}\ket{\psi_{ij}}_{B},
\end{equation}
where $\ket{\psi_{ij}}_{B}$ is the state of the black hole system $B$ with the ETW brane chosen to be of sub-flavours $i$ and $j$ respectively. The state can be prepared using a Euclidean path integral, and the Lorentzian description is obtained by analytic continuation as shown in \figref{fig:west_coast_intro}. 

A diagrammatic description of the boundary conditions that compute the overlap between two such states is
\begin{equation}\label{eq:overlap}
	\langle \psi_{i_1\,j_1}|\psi_{i_2\,j_2}\rangle = \ \includegraphics[scale = .7,valign = c]{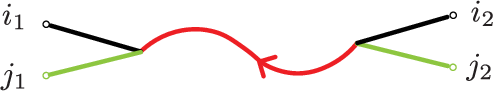},
\end{equation}
where the red line represents an asymptotic AdS boundary, whereas the black and green lines correspond to sub-flavour indices of the ETW brane.\footnote{Note that we have changed the diagrammatic notation from Ref.~\cite{Penington:2019kki}.} In this paper, we will consider two choices of boundary conditions imposed at the asymptotic boundary: a) microcanonical, where we impose a fixed energy $E$, and b) canonical, where we fix the renormalized length $\beta$ corresponding to the inverse temperature. The computation in \Eqref{eq:overlap} is done by performing the gravitational path integral over geometries consistent with these boundary conditions, e.g., 
\begin{equation}
	\langle \psi_{i_1\,j_1}|\psi_{i_2\,j_2}\rangle = \ \includegraphics[scale = .7,valign = c]{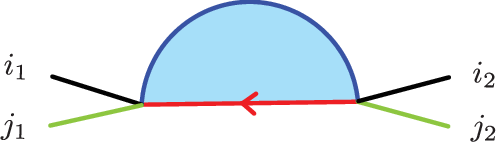},
\end{equation}
where the ETW brane has a definite flavour and thus, makes the diagram proportional to $\delta_{i_1,i_2}\delta_{j_1,j_2}$.


\subsection{Reflected Entropy} 
\label{sub:sr}

The reflected entropy $S_R(A:B)$ is a function defined for a density matrix $\rho_{AB}$ on a bipartite quantum system $AB$. One first considers the \emph{canonical purification} of $\rho_{AB}$ on a doubled Hilbert space 
\begin{equation}
	\ket{\sqrt{\rho_{AB}}} \in \mathrm{End}(\mathcal{H}_A) \otimes \mathrm{End}(\mathcal{H}_B) = (\mathcal{H}_A \otimes \mathcal{H}^*_A) \otimes (\mathcal{H}_B \otimes \mathcal{H}^*_B)~, 
\end{equation}
where the space of linear maps $\mathrm{End}(\mathcal{H}_A)$ acting on $\mathcal{H}_A$ itself forms a Hilbert space with inner product $\braket{X|Y} = \tr_A(X^\dagger Y)$. In general $\mathrm{End}(\mathcal{H})$ is isomorphic to the doubled copy $\mathcal{H} \otimes \mathcal{H}^*$. In other words, define $\ket{\sqrt{\rho_{AB}}}$ by finding the unique positive matrix square root of $\rho_{AB}$ and interpret the result as a state in $\mathrm{End}(\mathcal{H}_{A} \otimes \mathcal{H}_{B})$. This procedure is a generalization of the much more familiar procedure of purifying the thermal density matrix by considering the thermofield double state, discussed in the context of holography by Ref.~\cite{Maldacena:2001kr}.

The reflected entropy is then defined as \cite{Dutta:2019gen}: 
\begin{equation}
	S_R(A:B) = S(AA^*)_{\sqrt{\rho_{AB}}}.
\end{equation}
In fact, it has a two parameter Renyi generalization based on the following state:\footnote{This state is unnormalized as written. One can insert a factor of $({\rm Tr} \rho_{AB}^m)^{-1/2}$ to normalize it, though this is irrelevant in the $m\rightarrow 1$ limit.}
\begin{equation}\label{eq:psim} 
	\left| \psi^{(m)} \right> = \left| \rho_{AB}^{m/2} \right>. 
\end{equation}
The Renyi generalization, which we call the $(m,n)$-Renyi reflected entropy, is then given by 
\begin{equation}\label{eq:renyimn} 
	S_R^{(m,n)}(A:B) = - \frac{1}{n-1} \ln {\rm Tr} (\rho_{AA^\star}^{(m)})^n \, , \qquad \rho_{AA^*}^{(m)} = \frac{1}{ {\rm Tr} \rho_{AB}^m}{\rm Tr}_{BB^\star} \left| \rho_{AB}^{m/2} \right> \left< \rho_{AB}^{m/2} \right|, 
\end{equation}
for $m \geq 0, n \geq 0$.

In the West Coast Model, we are interested in calculating the reflected entropy $S_R(R_1:R_2)$. To motivate the proposed holographic answer, we can first consider preparing the Renyi generalization of the canonically purified state given in \Eqref{eq:psim}. For even integer $m$, this can be analyzed using the replica trick. Consider the Euclidean path integral that computes the norm of the state $\langle \psi^{(m)}|\psi^{(m)}\rangle = \tr(\rho_{R_1 R_2}^m)$, it is diagramatically represented by the boundary conditions: (e.g. $m=4$)
\begin{equation}
	 \tr(\rho_{R_1 R_2}^m) = \includegraphics[scale = .4,valign = c]{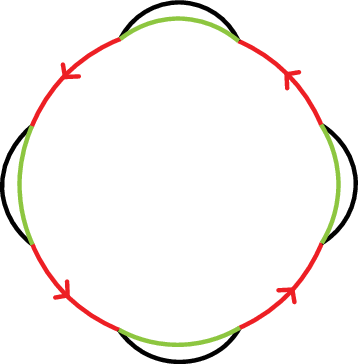}.
\end{equation}
The boundary conditions have a dihedral symmetry composed of $\mathbb{Z}_m$ rotations and a $\mathbb{Z_2}$ reflection. 

We now consider two natural saddle points that respect this replica symmetry \cite{Dutta:2019gen}, and applying the saddle point approximation gives us
\begin{align}
	\tr(\rho_{R_1 R_2}^m) &= \max \left[ \includegraphics[scale = .4,valign = c]{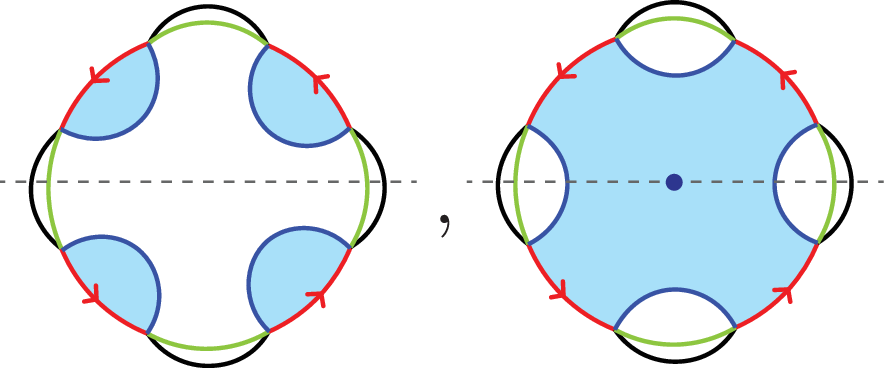} \right],\\
	&= \max \left[k Z_1^m,k^{m-1} Z_m \right],
\end{align}
where $Z_p$ is the $p$-boundary partition function and the purple dot indicates the location of the extremal surface. The disconnected saddle dominates before the Page transition, i.e. $k\ll \exp(S_0)$, while the connected saddle dominates after the Page transition. Each of these saddles has a $\mathbb{Z_2}$ symmetric slice which can be used to analytically continue the Euclidean saddle to a Lorentzian solution, giving us two candidate geometries corresponding to $\ket{\psi^{(m)}}$:
\begin{equation}\label{eq:cp_geom}
	\ket{\psi^{(m)}}= \includegraphics[scale = .4,valign = c]{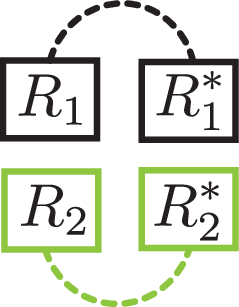} \qquad \text{or} \qquad  \includegraphics[scale = .4,valign = c]{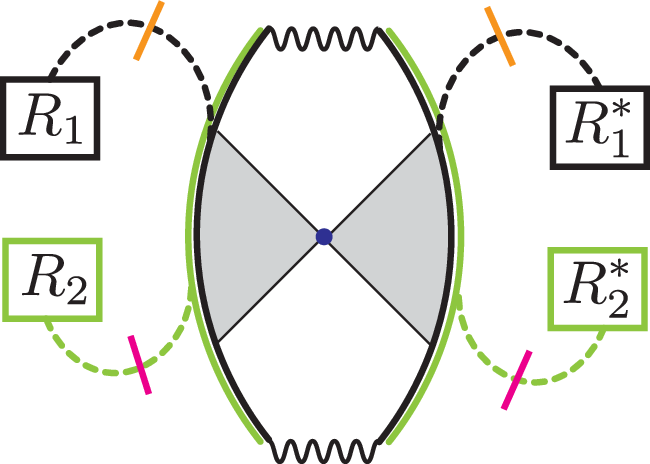}.
\end{equation}
We refer to these geometries as the disconnected and connected purifications respectively. Here, $m$ is simply a parameter that fixes the area of the extremal surface and thus, the analytic continuation in $m$ is straightforward. As $m\rightarrow 1$, these geometries can be interpreted as arising from gluing together two copies of the entanglement wedge of $R_1\cup R_2$ \cite{Engelhardt:2017aux,Engelhardt:2018kcs}.\footnote{See also Ref.~\cite{Engelhardt:2022qts} for a  construction of the canonical purification of the black hole system in a model of JT gravity coupled to matter.}

Within the respective phases, we can apply the QES formula once we know the geometry corresponding to $\ket{\psi^{(m)}}$. For the disconnected saddle, the QES is trivial and thus, the reflected entropy vanishes. On the other hand, for the connected saddle we have two non-trivial candidate QESs represented by the pink and orange cuts in \Eqref{eq:cp_geom}. This leads to the proposed formula 
\begin{equation}\label{eq:wcsr2}
	S_R(R_1:R_2) = \begin{cases}
		0 & k < \exp(S_{BH})\\
		2 \min(\log k_1,\log k_2) &  k>\exp(S_{BH})
	\end{cases}
\end{equation}
While this saddle point approximation is expected to work far away from the Page transition, there are various issues with this proposal which will require a more detailed analysis that we carry out in \secref{sec:SR}. In particular, the saddle point calculation suffers from an order of limits issue when analytically continuing $m,n\rightarrow 1$. Further, \Eqref{eq:wcsr2} has a discontinuous jump at the phase transition which is resolved by non-perturbative effects. Once these effects are included, we will find significant corrections to this formula near the phase transition.


\section{Entanglement Spectrum} 
\label{sec:EE}

Before we go to a more careful analysis of the reflected entropy, we first discuss the simpler case of entanglement entropy. In \secref{sub:resolvent_ee}, we first review the resolvent trick that was used to compute the entanglement spectrum in Ref.~\cite{Penington:2019kki}. This serves as a warm-up for the analogous calculation for the reflected spectrum. Further, we analyze the Renyi entropies in the West Coast Model in \secref{sub:renyi} which will be useful for our later calculations.

\subsection{Resolvent Trick} 
\label{sub:resolvent_ee}

The technique we use here was presented in Ref.~\cite{Penington:2019kki} (see also \cite{Akers:2020pmf,Shapourian:2020mkc,Dong:2021oad}). This powerful approach enables us to write down a Schwinger-Dyson equation for the \textit{resolvent} of $\rho_{R_1 R_2}$, which then gives full information about the entanglement spectrum.

Consider the resolvent matrix $R_{ij}(\lambda)$ for the density matrix $\rho_{R_1 R_2}$ defined formally as
\begin{align}
    R_{ij}(\lambda)=\left( \frac{1}{\lambda-\rho_{R_1 R_2}}  \right)_{ij},
\end{align}
where the $i,j$ indices run over both $R_1$ and $R_2$ labels and take values from $1$ to $k=k_1 k_2$. We then define the resolvent as the trace $R(\lambda)=\tr R_{ij}(\lambda)$. From the resolvent, we can obtain the density of eigenvalues $D(\lambda)$ using
\begin{align}
    D(\lambda) &= -\frac{1}{\pi}\lim_{\epsilon \rightarrow 0} \text{Im}\, R(\lambda+i\epsilon).
\end{align}
To evaluate $R(\lambda)$, we expand the matrix inverse around $\lambda=\infty$ as
\begin{equation}
  R(\lambda)=\frac{k}{\lambda}+\sum_{n=1}^{\infty} \frac{\tr (\rho_{R_1 R_2}^n)}{\lambda^{n+1}},
\end{equation}
sum the series and then analytically continue to $\lambda \in [0,1]$ on the real axis where $D(\lambda)$ is non-zero. In terms of diagrams, this leads to the boundary conditions:
\begin{equation}
  \includegraphics[width=0.85\textwidth]{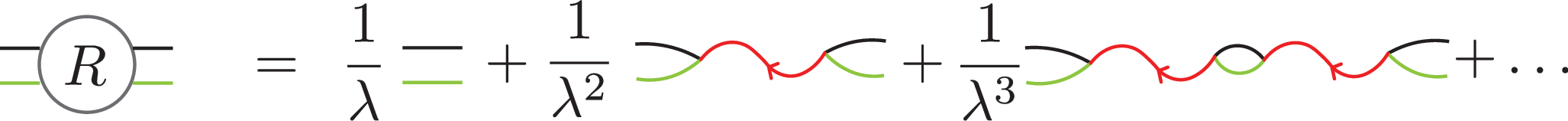},        
\end{equation}
where uncontracted indices have been used to represent a matrix equation for $R_{ij}$ and each red line carries a power of $\frac{1}{k Z_1}$ for normalization. These boundary conditions are then filled in with planar disk geometries since crossing (higher genus) geometries are suppressed by powers of $\frac{1}{k}$ ($e^{-S_0}$). 

Now denote $F_{ij}(\lambda)$ to be the connected part of the resolvent, defined by
\begin{equation}
  \label{eq:resolv2}
  \includegraphics[width=0.85\textwidth]{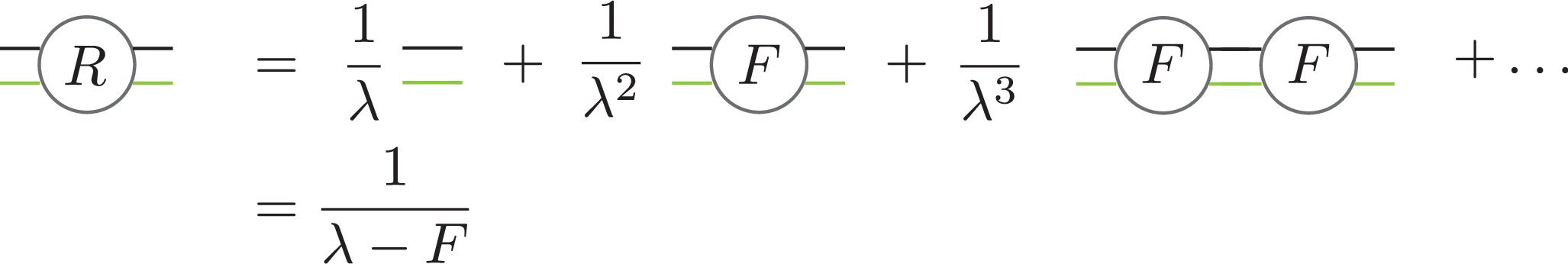},      
\end{equation}
where the last term denotes the matrix $(\lambda-F)^{-1}$. We can now write down a Schwinger-Dyson equation for $R_{ij}(\lambda)$ and $F_{ij}(\lambda)$:
\begin{equation}\label{eq:F}
  \includegraphics[width=0.85\textwidth]{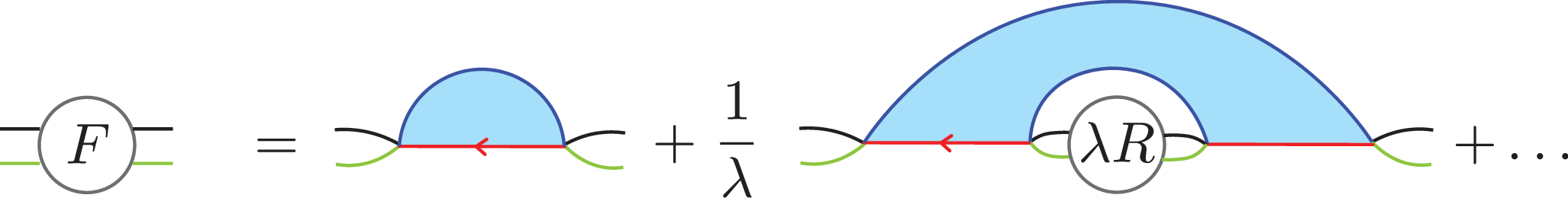},     
\end{equation}
where we note that a self consistent solution is to consider every term to be proportional to $\delta_{ij}$, and thus $F_{ij}(\lambda)=\frac{F(\lambda)}{k}\delta_{ij}$, where we have defined $F(\lambda)=\tr F_{ij}(\lambda)$. After taking the trace of \Eqref{eq:F}, we obtain the following algebraic equation
\begin{align}\label{eq:Fsum}
  F(\lambda) &= \sum^{\infty}_{n=1} \frac{k Z_n R(\lambda)^{n-1}}{(k Z_1)^n}.
\end{align}
We could now consider boundary conditions that correspond to either the microcanonical or canonical ensemble. The microcanonical case is identical to a random state and is discussed in detail in Ref.~\cite{Akers:2021pvd},\footnote{In this model, fixing the energy is the same as fixing the area of the horizon in the semiclassical approximation. Thus, the calculations can be understood as related to those done in random tensor networks via the connection to fixed-area states \cite{Akers:2018fow,Dong:2018seb,Dong:2019piw}} so our primary focus here will be on the canonical ensemble. For this case, we can use an integral representation for the $n$-boundary partition function $Z_n$ given by
\begin{equation}
	Z_{n}=\int_0^\infty ds\,\rho(s)y(s)^n,
\end{equation}
where $\rho(s)$ is the density of states, and $y(s)$ is the Boltzmann weight for the thermal spectrum given by
\begin{align}
	\rho(s)&=\frac{e^{S_0}s \sinh{2\pi s}}{2 \pi^2}\\
	y(s)&=e^{-\frac{\beta s^2}{2}} 2^{1-2 \mu} |\Gamma(\mu-\frac{1}{2}+i s)|^2.
\end{align}
It is convenient to define the normalized Boltzmann weight $w(s)=\frac{y(s)}{Z_1}$. Using this representation in \Eqref{eq:Fsum}, we obtain
\begin{equation}
	F(\lambda) = \int_{0}^{\infty} ds \frac{\rho(s) w(s)}{k-w(s)R}.
\end{equation}
Finally, combining this with \Eqref{eq:resolv2}, we obtain an integral equation for $R(\lambda)$:
\begin{align}
 \lambda R = k+\int_{0}^{\infty} ds \frac{\rho(s) w(s) R}{k-w(s)R}
\end{align}
Solving this equation will give us the entanglement spectrum and corresponding entropies.

\subsubsection*{Approximate Entanglement Spectrum}

\begin{figure}[h!]
    \centering \includegraphics[width=0.8\textwidth]{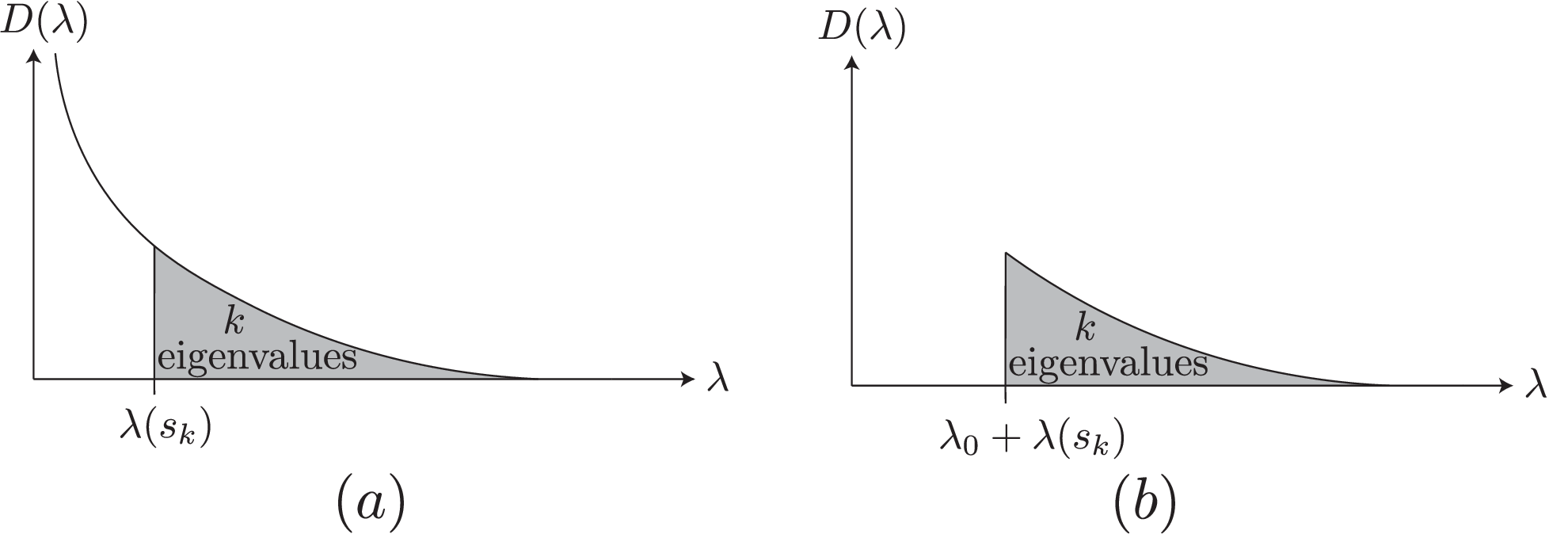}
    \caption{(a) The thermal spectrum of the JT black hole. (b) The approximate spectrum of $\rho_{R_1 R_2}$ is a cutoff thermal spectrum which is obtained by truncating the spectrum to the $k$ largest eigenvalues and shifting it by $\lambda_0$ to make it normalized.}
    \label{fig:cutoff}
\end{figure}

The resolvent equation was solved approximately in Ref.~\cite{Penington:2019kki} to obtain an approximate entanglement spectrum. It was shown that the spectrum of $\rho_{R_1 R_2}$ can be well approximated by a cutoff thermal spectrum (see \figref{fig:cutoff}), i.e.,
\begin{equation}\label{eq:cutoff}
	D(\lambda) = \int_{0}^{s_k} ds\,\rho(s)\delta(\lambda-\lambda_0-w(s)),
\end{equation}
where the parameters $s_k$ and $\lambda_0$ can be computed by
\begin{align}\label{eq:sk}
 k &= \int_{0}^{s_k} ds\, \rho(s)\\
	\lambda_0 &= \frac{1}{k}\int_{s_k}^{\infty} \rho(s) w(s).\label{eq:l0}
\end{align}
For more details about the calculation, refer to Appendix F of Ref.~\cite{Penington:2019kki}. Here, we will simply use the result and discuss the interpretation of this spectrum. 

Intuitively, the thermal spectrum of the black hole is divided into two portions - a part that has already undergone the Page transition, and a part that hasn't. For states corresponding to sufficiently high energy in the thermal spectrum, the Page transition has still not occurred since they correspond to larger black holes. On the other hand, low energy states are past the Page transition and in this part of the wavefunction, the radiation system contains an island in the black hole interior. Thus, it is natural to expect that details of the spectrum such as the probability distribution of the horizon area should be observable. This natural ansatz is in fact a good approximation and leads to a cutoff thermal spectrum where $s_k$ roughly corresponds to the threshold dividing the pre-Page and post-Page parts of the wavefunction. The post-Page portion of the spectrum is not normalized on its own and thus, $\lambda_0$ is the additive constant contributed by the pre-Page spectrum that shifts the spectrum of $\rho_{R_1 R_2}$ to make it appropriately normalized. Further, this ansatz captures the feature that the rank of $\rho_{R_1 R_2}$ cannot exceed $k$ even far past the Page transition, since the dimension of the radiation Hilbert space is $k$.


\subsection{Renyi Entropy} 
\label{sub:renyi}

Given this approximate description of the spectrum, we will now analyze the Renyi entropies in this model,
\begin{equation}
	S_m(\rho_{R_1 R_2}) = -\frac{1}{m-1} \log (\tr (\rho_{R_1 R_2}^m))
\end{equation}
for different values of $m$. To simplify the analysis, we choose $\mu \gg \frac{1}{\beta}$. In this limit, the $\mu$-dependent terms drop out from $w(s)$. Further, we also work in the semiclassical limit $\beta \rightarrow 0$.\footnote{Note that $\beta$ should be replaced with $\beta G_N$ to restore explicit dependence on the Newton constant.}

Using \Eqref{eq:cutoff}, we obtain
\begin{equation}\label{eq:mth_moment}
	\tr (\rho_{R_1 R_2}^m) = \int_{0}^{s_k} \rho(s) (\lambda_0+w(s))^m.
\end{equation}
There are a few values of $k$ near the phase transition, equivalently $s_k \approx \frac{\log k-S_0}{2\pi}$, that play an important role in this calculation. First, the location of the peak of $\rho(s) w(s)^m$ is $s_m=\frac{2\pi}{m\beta}$. Secondly, far from the phase transition the calculation of Renyi entropies is dominated by either of the two $\mathbb{Z}_m$ symmetric saddles. The location at which they exchange dominance is denoted by $s_k^{(m)}$. Working in the saddle point approximation, we obtain the condition
\begin{equation}\label{eq:skm}
	\frac{1}{k^{m-1}} = \exp\left[(1-m)S_0+\frac{2\pi^2}{\beta}(\frac{1}{m}-m)\right],
\end{equation}
which leads to $s_k^{(m)}\approx \frac{\pi}{\beta}(1+\frac{1}{m})$. Lastly, from \Eqref{eq:l0}, we see that $\lambda_0$ has a phase transition at $s_k=s_1$.

With these ingredients in hand, we can now analyze \Eqref{eq:mth_moment}. An important feature that appears in this model is that different Renyi entropies have transitions at different values of $k$. We will analyze them separately for $m>1$, $m=1$ and $m<1$ since they have qualitatively different behaviour.

\begin{figure}
	\centering 
	\begin{subfigure}
		[b]{0.47
		\textwidth} 
		\includegraphics[width=
		\textwidth]{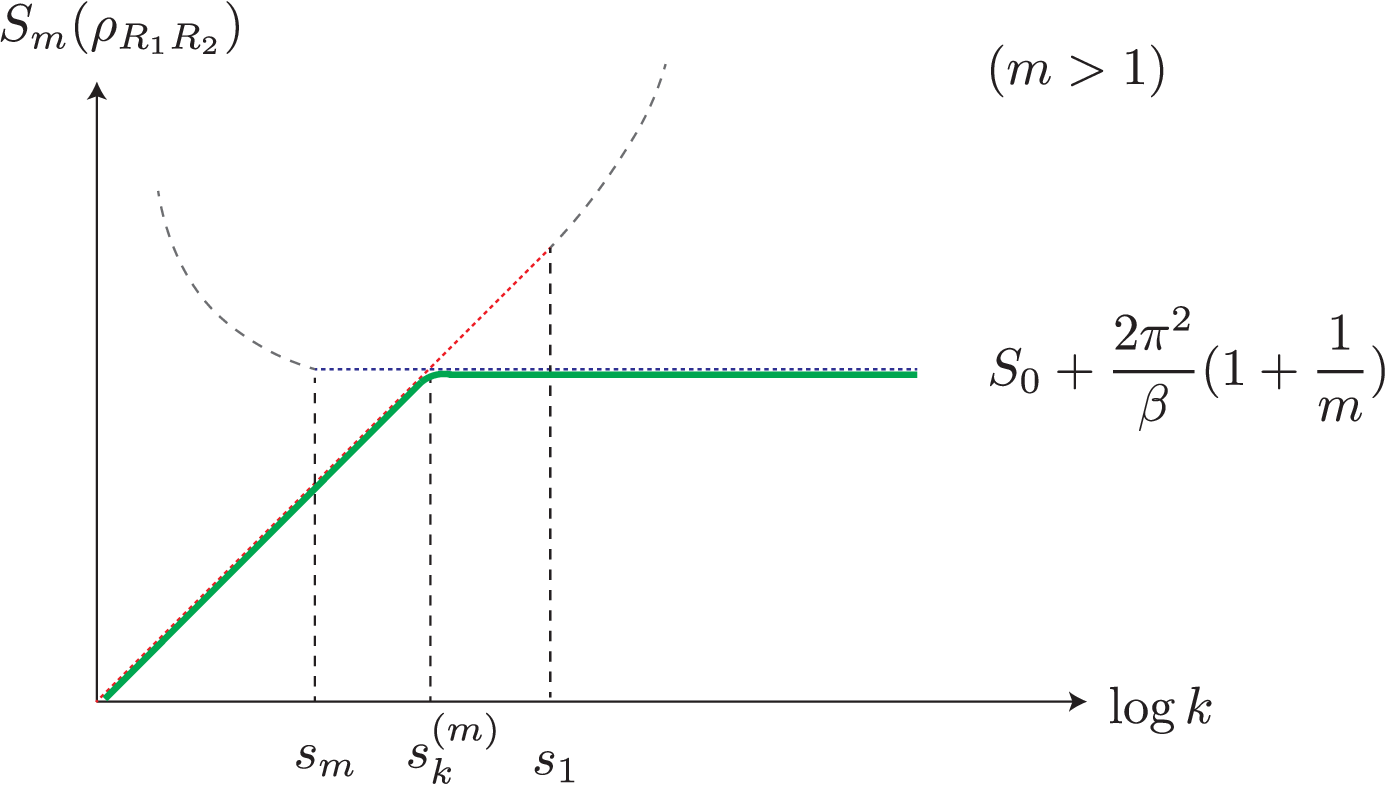} \subcaption{} 
	\end{subfigure}
	\hspace{5mm} 
	\begin{subfigure}
		[b]{0.47
		\textwidth} 
		\includegraphics[width=
		\textwidth]{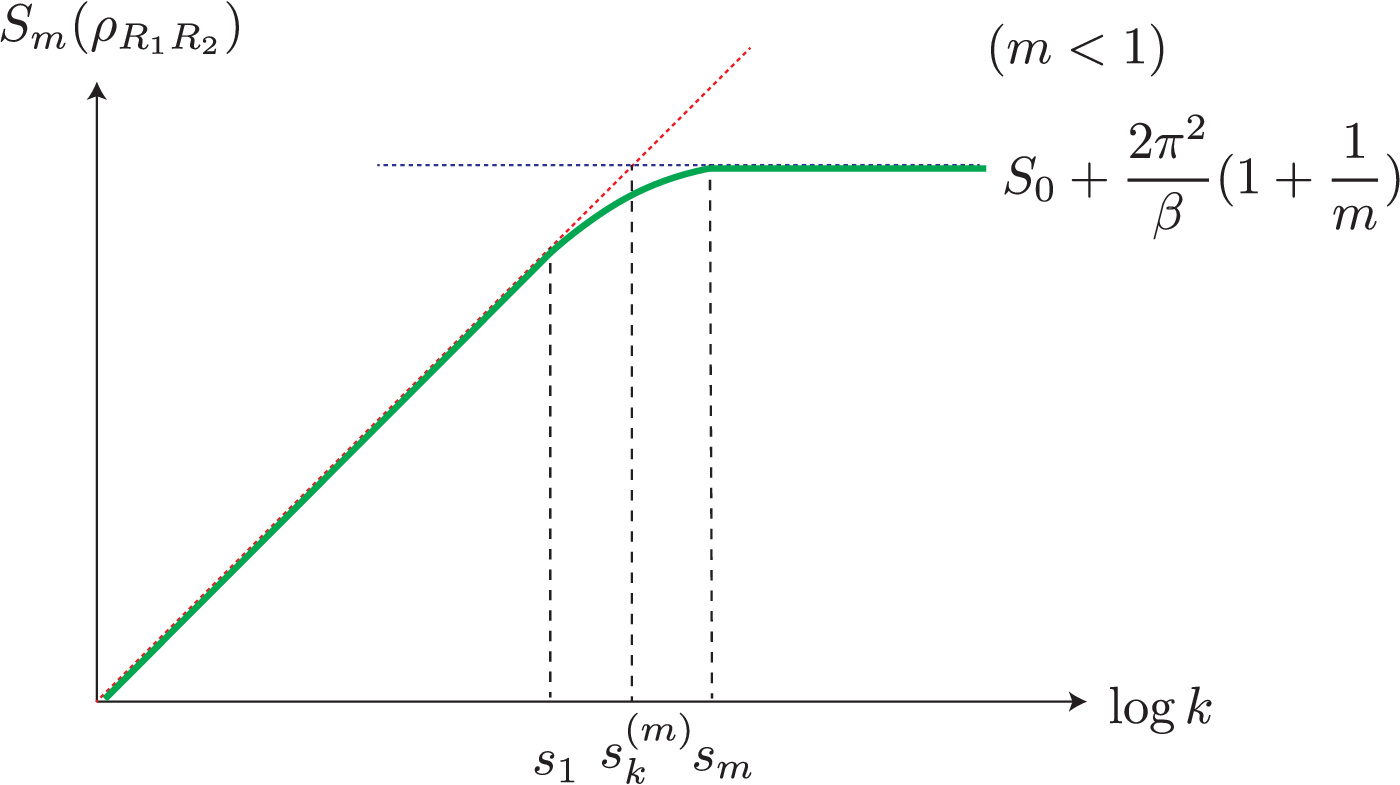} \subcaption{} 
	\end{subfigure}
	\caption{(a) For $m>1$, the Renyi entropy (denoted by the green line) has a sharp transition at $s_k=s_k^{(m)}$ where the two replica symmetric saddle contributions (denoted by blue and red dotted lines) exchange dominance. Phase transitions in the two terms in \Eqref{eq:max_m} happen at $s_k=s_1,s_m$ (depicted with curved, gray dashed lines) and thus don't affect the Renyi entropies. (b) For $m<1$, the Renyi entropy (denoted by the green line) has a transition in the range $s_k \in [s_1,s_m]$ whereas the two naive replica symmetric saddles (denoted by blue and red dotted lines) exchange dominance at $s_k=s_k^{(m)}$. Thus, the phase transition is over a large window of size $O(\frac{1}{\beta})$ and the Renyi entropy has $O(\frac{1}{\beta})$ corrections compared to a naive analytic continuation of the replica symmetric saddles.} 
\label{fig:renyis} \end{figure}

\subsubsection*{\underline{$m>1$}}

For $m>1$, we have $s_m<s_k^{(m)}<s_1$. This implies that $\lambda_0 \approx \frac{1}{k}$ for $s_k \approx s_k^{(m)}$. As a simple approximation, we can compute \Eqref{eq:mth_moment} by
\begin{equation}\label{eq:max_m}
	\tr(\rho_{R_1 R_2}^m) \approx \max \left[\int_{0}^{s_k}ds\, \rho(s)\lambda_0^m,\int_{0}^{s_k}ds\, \rho(s) w(s)^m\right].
\end{equation}
Since at the level of the saddle point approximation, this comparison is identical to \Eqref{eq:skm}, we see that $\tr(\rho_{R_1 R_2}^m)$ has a phase transition at $s_k=s_k^{(m)}$. More so, the second term in \Eqref{eq:max_m} is the integral of a Gaussian with a width $\sigma_m=\frac{1}{\sqrt{m\beta}}$. Since $s_k^{(m)}-s_m \gg \sigma_m$, the second term is almost unchanging in this region. Thus, we see that for $m>1$, $\tr(\rho_{R_1 R_2}^m)$ has a rather sharp phase transition. The Renyi entropy takes the form
\begin{equation}
	S_m(\rho_{R_1 R_2}) = \min \left[ \log k, S_0 + \frac{2\pi^2}{\beta }(1+\frac{1}{m}) \right],
\end{equation}
as shown in \figref{fig:renyis}.

\subsubsection*{\underline{$m=1$}}

The case $m=1$ corresponds to the entanglement entropy which was analyzed in detail in Ref.~\cite{Penington:2019kki}. In this case, $s_m=s_k^{(m)}=s_1$ and thus, there isn't a sharp transition. Instead, one finds enhanced corrections of $O(\frac{1}{\sqrt{\beta}})$ in a window of size $\Delta \log k =O(\frac{1}{\sqrt{\beta}})$ near $s=s_1$. Again, we can use the approximate spectrum in \Eqref{eq:cutoff} to obtain
\begin{equation}
	S(\rho_{R_1 R_2}) \approx -\int_{0}^{s_k} \rho(s)(\lambda_0+w(s))\log(\lambda_0 +w(s)).
\end{equation}

\subsubsection*{\underline{$m<1$}}

For $m<1$, we instead have $s_m>s_k^{(m)}>s_1$. Thus, we see that the phase transition in $\tr(\rho_{R_1 R_2}^m)$ starts happening at $s_k=s_1$ when $\lambda_0$ starts decaying. This is qualitatively different from the $m>1$ case and will require a more careful analysis.\footnote{Similar results have been discussed in Ref.~\cite{Dong:2021oad}. We are very grateful to Sean Mcbride and Wayne Weng for discussions related to this topic.}

In order to compute \Eqref{eq:mth_moment}, we can use the following inequality on the integrand:
\begin{equation}\label{eq:bound}
	\max[\lambda_0^m,w(s)^m] \leq \left(\lambda_0+w(s)\right)^m \leq \lambda_0^m+w(s)^m,
\end{equation}
where the lower bound follows from positivity, while the upper bound follows from concavity. Upon performing the integral for $s_k = s_1 - O(\frac{1}{\beta})$, since $\lambda_0 \gg w(s)$, we obtain
\begin{equation}
	\tr(\rho_{R_1 R_2}^m) \approx k \lambda_0^m \approx \frac{1}{k^{m-1}},
\end{equation}
since both the upper and lower bounds take the same value at leading order. In the regime $s_1+O(\frac{1}{\beta})<s_k<s_m-O(\frac{1}{\beta})$, both the terms are comparable. Thus, we obtain
\begin{equation}
	\tr(\rho_{R_1 R_2}^m) \approx \# \exp \left[2\pi s_k-\frac{m \beta s_k^2}{2}+(1-m) S_0-\frac{2\pi^2 m}{\beta}\right],
\end{equation}
where there is a potential $O(1)$ multiplicative uncertainty. Similarly in the regime $s_k=s_m+O(\frac{1}{\beta})$, the $w(s)$ term dominates over the $\lambda_0$ term giving us
\begin{equation}
	\tr(\rho_{R_1 R_2}^m) \approx 		\exp \left[(1-m)S_0+\frac{2\pi^2}{\beta}\left(\frac{1}{m}-m \right)\right],
\end{equation}
where we have used the saddle point approximation. Combining these results, we obtain
\begin{equation}
	S_m(\rho_{R_1 R_2})\approx \begin{cases}
		\log k & s_k < s_1\\
		\frac{2\pi s_k-\frac{m \beta s_k^2}{2}+(1-m) S_0-\frac{2\pi^2 m}{\beta}}{1-m} & s_1 < s_k < s_m\\
		S_0+\frac{2\pi^2}{\beta}\left(1+\frac{1}{m}\right) &  s_k > s_m
	\end{cases},
\end{equation}
where we have ignored subleading corrections at $s_k=s_1$ and $s_k=s_m$. The result is sketched in \figref{fig:renyis}, along with the naive expectation from analytically continuing the results of the $\mathbb{Z}_m$ symmetric saddles a la Lewkowycz-Maldacena \cite{Lewkowycz:2013nqa,Dong:2016fnf}. We see that there are in fact large $O(\frac{1}{\beta})$ corrections to the naive holographic answer for Renyi index $m<1$. This is consistent with expectations, hinted at in Ref.~\cite{Murthy:2019qvb}, based on chaotic behaviour. As shown, the transition happens over a parametrically large window of $\Delta \log k \approx O(\frac{1}{\beta})$.

Before moving on, we analyze the problem a little more carefully in a window $s_k=s_1 \pm O(\frac{1}{\sqrt{\beta}})$ since this will be relevant for the reflected entropy transition. We can use the inequality in \Eqref{eq:bound} and perform the integrals, which are simple Gaussian integrals. In fact, we will include one-loop effects to find the answer to greater accuracy. Computing the relevant quantities, we obtain
\begin{align}\label{eq:gaussian}
\begin{split}
    Z_1 &\simeq \frac{e^{S_0+2\pi^2/\beta}}{\sqrt{2\pi\beta^3}} \\
    k & \simeq e^{S_0}\frac{s_k e^{2\pi s_k}}{8\pi^3} \\
    \lambda_0 &\simeq \frac{\text{erfc}(\frac{\beta s_k-2\pi}{\sqrt{2\beta}})}{2k} \\
    \int_0^{s_1} \rho(s)w(s)^m &\simeq  \frac{\beta^{3m/2-1}e^{4\pi^2(1-m)/\beta}}{(1-m)(2\pi)^{m/2-2}},
    \end{split}
\end{align}
where $\text{erfc}(x)=\frac{2}{\sqrt{\pi}}\int_x^\infty e^{-x^2}dx$ is the error function. Thus, integrating \Eqref{eq:bound} with the measure $\rho(s)$ and using \Eqref{eq:gaussian}, we obtain
\begin{equation}\label{eq:prob}
  \tr(\rho_{R_1 R_2}^m) \approx k \lambda_0^m,
\end{equation}
where we have used the fact that the integral of $\rho(s)w(s)^m$ is suppressed by powers of $\beta$ in the regime of interest. 



\section{Reflected Entanglement Spectrum} 
\label{sec:SR}

We now use a generalization of the resolvent trick in \secref{sub:resolvent_sr} to analyze the reflected entanglement spectrum. Solving the Schwinger-Dyson equation leads us to the spectrum and reflected entropy in \secref{sub:spectrum}. Finally, we analyze the $(m,n)$-Renyi reflected entropies in \secref{sub:renyi_sr}.

\subsection{Resolvent Trick} 
\label{sub:resolvent_sr}

Our goal here is to find the resolvent for the reduced density matrix $\rho^{(m)}_{R_1 R_1^*}$ obtained from the Renyi generalization of the canonically purified state $\ket{\psi^{(m)}}$ defined in \Eqref{eq:psim}, i.e., 
\begin{equation}
	R_m(\lambda) = \frac{k_1^2}{\lambda} + \sum^\infty_{n=1} \frac{\tr (\rho^{(m)}_{R_1 R_1^*})^n}{\lambda^{n+1}} 
\end{equation}
where the integer moments of the (normalized) density matrix can be computed using the replica trick on $m n$ copies of the system \cite{Dutta:2019gen,Akers:2021pvd}. 

In order to do so, we first set up a slightly more general problem.\footnote{This problem is analyzed in more detail in Ref.~\cite{Akers:2021pvd}, here we attempt to be brief and focus on the main results.} Consider the following $2 \times 2$ ``matrix'' of resolvents:
\begin{equation}\label{eq:res1}
	\mathbb{R}(\lambda) =   \sum_{k=0}^\infty \lambda^{-1-k/2}  \begin{pmatrix} 0 & (\rho^{m/2}_{R_1 R_2})^{\Gamma^\dagger}   \\  (\rho^{m/2}_{R_1 R_2})^\Gamma  &  0 \end{pmatrix}^k,
\end{equation}
where $\rho \rightarrow \rho^\Gamma$ is an involution defined to take a linear operator $\rho$ on $\mathcal{H}_{R_1 R_2}$ to a linear operator from $\mathcal{H}_{R_1 R_1^\star} \rightarrow \mathcal{H}_{R_2 R_2^\star}$. It is defined by re-arranging the incoming/outgoing legs in the obvious (and canonical/basis independent) way. Each insertion of $(\rho^{m/2}_{R_1 R_2})^{\Gamma}$ (and $(\rho^{m/2}_{R_1 R_2})^{\Gamma^\dagger}$) has $m/2$ replicas of each of the bra and ket of the original state $\ket{\psi}$. These involuted density operators are related to the density matrices in the canonically purified state by 
\begin{equation}
	(\rho^{m/2}_{R_1 R_2})^{\Gamma^\dagger} (\rho^{m/2}_{R_1 R_2})^\Gamma = \rho^{(m)}_{R_1 R_1^\star} \,, \qquad (\rho^{m/2}_{R_1 R_2})^\Gamma (\rho^{m/2}_{R_1 R_2})^{\Gamma^\dagger} = \rho^{(m)}_{R_2 R_2^\star} . 
\end{equation}
Diagramatically this is given by the following boundary conditions
\begin{equation}\label{eq:gamma}
	 (\rho^{m/2}_{R_1 R_2})^{\Gamma} = \includegraphics[scale = .4,valign = c]{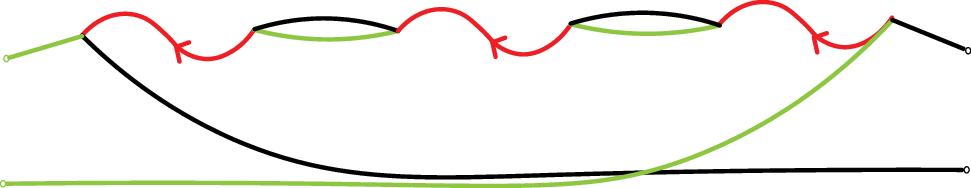},
\end{equation}
where we have chosen $m=6$ for example. As before, any calculation using the gravitational path integral in this model will amount to a sum over geometries of different topology connecting the various asymptotic boundaries, e.g, in computing \Eqref{eq:gamma} we get
\begin{equation}
	(\rho^{m/2}_{R_1 R_2})^{\Gamma} = \sum_{\text{topologies}}\includegraphics[scale = .5,valign = c]{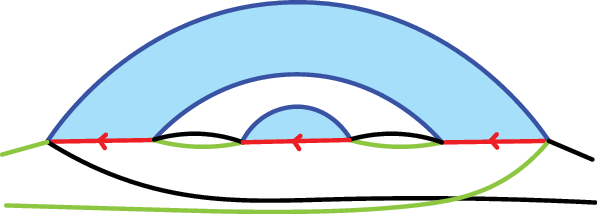}.
\end{equation}
For convenience, we will use a shorthand representation
\begin{equation}\label{eq:short}
 	\includegraphics[scale = .75,valign = c]{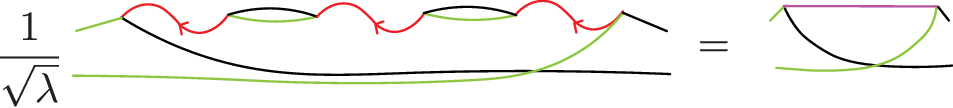}
\end{equation}

From these results we can infer an alternative expression for matrix $\mathbb{R}(\lambda)$: 
\begin{equation}
	\mathbb{R}(\lambda) = \sum_{n=0}^\infty \lambda^{-1-n} 
	\begin{pmatrix}
		(\rho^{(m)}_{R_1 R_1^*})^{n} & 0 \\
		0 & (\rho^{(m)}_{R_2 R_2^*})^{n} 
	\end{pmatrix}
	+\sum_{n=1}^\infty \lambda^{-1/2-n} 
	\begin{pmatrix}
		0 & \rho_{R_1 R_1^\star}^{n-1} (\rho^{m/2}_{R_1 R_2})^{\Gamma^\dagger} \\
		\rho_{R_2 R_2^\star}^{n-1} (\rho^{m/2}_{R_1 R_2})^\Gamma & 0 
	\end{pmatrix}
	, 
\end{equation}
which diagrammatically looks like 
\begin{equation}
	\includegraphics[width=.9
	\textwidth]{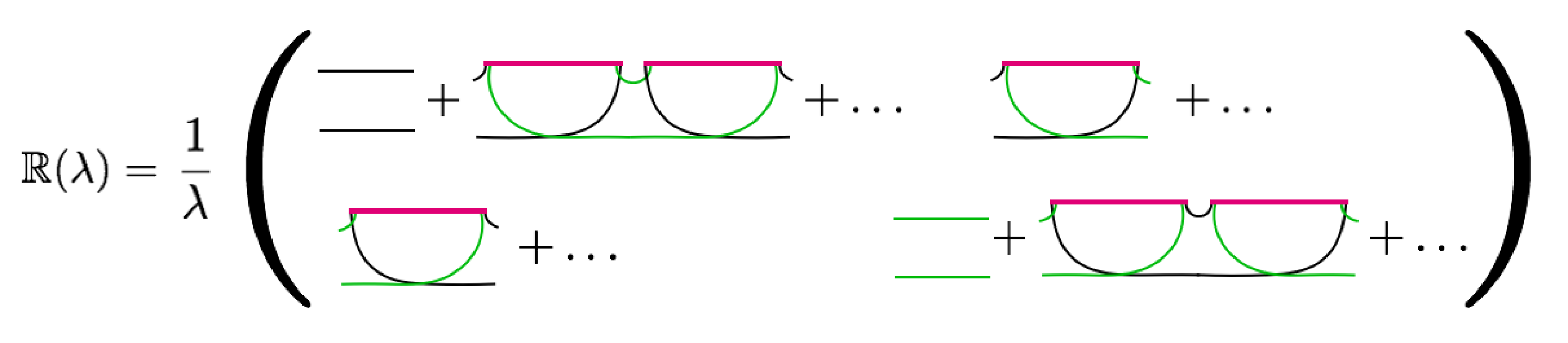}. 
\end{equation}
The resolvent for system $R_1\cup R_1^{*}$ in the state $\ket{\rho^{m/2}_{R_1\,R_2}}$ is given by
\begin{equation}
	R(\lambda) = {\rm Tr}_{R_1 R_1^\star} (\mathbb{R}_{11}(\lambda))
\end{equation}

\begin{figure}
    \centering
    \includegraphics[width=\textwidth]{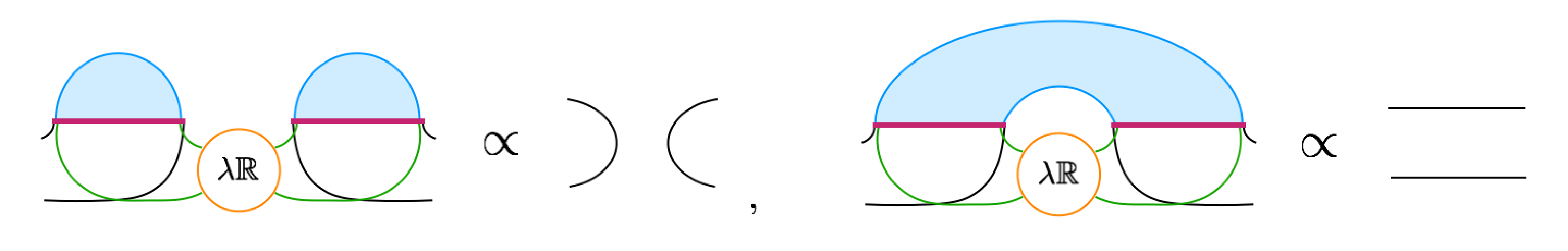}
    \caption{Example geometries that give rise to the projector or the identity. All possible contributions are either proportional to the identity operator $1_{R_1 R_1^*}, 1_{R_2 R_2^*}$, or one of the four projectors $\ket{\epsilon_{R_1,R_2}}\bra{\epsilon_{R_1,R_2}}$.}
    \label{fig:proj_id}
\end{figure}

A single copy of $(\rho^{(m/2)}_{R_1 R_2})^{\Gamma}$ is referred to as a \emph{puddle} and given two contiguous sets of puddles, there are two classes of diagrams, either \emph{disconnected} or \emph{connected}. As before, consider the connected part of $\mathbb{R}$ that we call $\mathbb{F}$. This corresponds to a sum over the sub-diagrams over all connected contractions . We have 
\begin{align}\label{eq:F_def} 
	\mathbb{R}(\lambda) = \frac{1}{\lambda} + \frac{\mathbb{F}}{\lambda^2} + \frac{\mathbb{F}^2}{\lambda^3} + \ldots = \frac{1}{\lambda-\mathbb{F}} 
\end{align}
A useful fact is that all diagrams in $\mathbb{F}$ and $\mathbb{R}$ take the following form: (see \figref{fig:proj_id})
\begin{align}\label{eq:rmatrix}
	\mathbb{F}(\lambda) &= 
	\begin{pmatrix}
		G_{11} (1_{R_1 R_1^\star} - e_{R_1}) + F_{11} e_{R_1} & F_{12} \left| \epsilon_{R_1} \right> \left< \epsilon_{R_2} \right| \\
		F_{21} \left| \epsilon_{R_2} \right> \left< \epsilon_{R_1} \right| & G_{22} (1_{R_2 R_2^\star} -e_{R_2}) +F_{22} e_{R_2} 
	\end{pmatrix}
	, \\
	\mathbb{R}(\lambda) &= 
	\begin{pmatrix}
		S_{11} (1_{R_1 R_1^\star} -e_{R_1}) + R_{11} e_{R_1} & R_{12} \left| \epsilon_{R_1} \right> \left< \epsilon_{R_2} \right| \\
		R_{21} \left| \epsilon_{R_2} \right> \left< \epsilon_{R_1} \right| & S_{22} (1_{R_2 R_2^\star}-e_{R_2}) +R_{22} e_{R_2} 
	\end{pmatrix}
	,
\end{align}
where $\ket{\epsilon_{R_1}} = \frac{1}{\sqrt{k_1}}\ket{1_{R_1}}$ is the maximally mixed state on $R_1 R_1^*$ (same for $\ket{\epsilon_{R_2}}$), $e_{R_1} = \left| \epsilon_{R_1} \right> \left< \epsilon_{R_1} \right|$ and $e_{R_2} = \left| \epsilon_{R_2} \right> \left< \epsilon_{R_2} \right|$ are normalized minimal projectors and $F$ is a $2\times 2$ matrix of scalars (not to be confused with $\mathbb{F}$). 

In order to write down a Schwinger-Dyson equation for $\mathbb{F}(\lambda)$, we can organize the different diagrammatic contributions to $\mathbb{F}$ by the number of $\mathbb{R}$ insertions, labelled $k-1$, where $k$ is the number of puddles in each diagram: 
\begin{equation}\label{eq:F_as_R}
	\includegraphics[width=.85
	\textwidth]{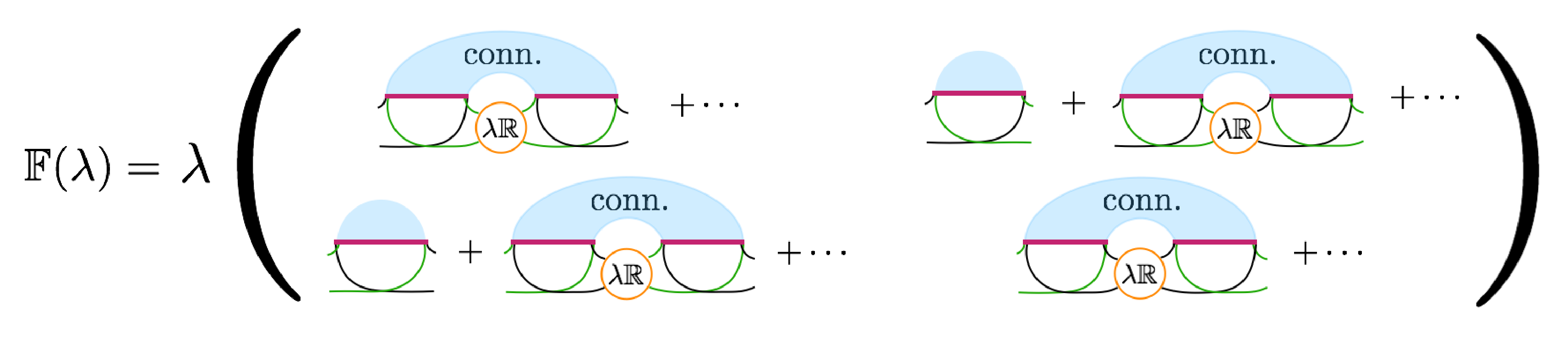} 
\end{equation}
where the shorthand \textit{conn.} indicates connected contractions. In the above figure, the bubbles of $\lambda \mathbb{R}$ indicate insertions of the resolvent matrix given in \Eqref{eq:rmatrix} and the various components of the matrix are represented by coloured index lines. Thus, combining \Eqref{eq:F_as_R} and \Eqref{eq:F_def}, we now have self-consistency equations that we can solve to obtain $\mathbb{R}$ and $\mathbb{F}$, order by order in $k$.

\subsection{Spectrum and Reflected Entropy} 
\label{sub:spectrum}

The solution to the Schwinger-Dyson equation is identical to that obtained in Ref.~\cite{Akers:2021pvd} for a random tripartite state. Thus, we only discuss the results here. The results follow from a conjecture that the important contributions to the matrix $\mathbb{F}$ come from the lowest two orders in $k$, i.e. $\mathbb{F}\simeq \mathbb{F}^{(1)}+\mathbb{F}^{(2)}$, in the limit we are considering. Evidence for this conjecture in the case of the microcanonical ensemble is provided in Ref.~\cite{Akers:2021pvd}, while here we simply apply the same conjecture, now to the case of the canonical ensemble, and obtain physically sensible results. 

Since the solution to the Schwinger-Dyson equation is obtained by considering $k=1$ and $k=2$, the two parameters that determine the spectrum turn out to be given by (e.g. for $m=6$)
\begin{align}\label{eq:bmdm}
	D_m&=\frac{\sqrt{\lambda}}{\sqrt{k_1 k_2}}\,\,\includegraphics[scale = .5,valign = c]{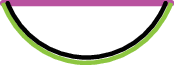} = \frac{\sqrt{\lambda}}{\sqrt{k_1 k_2}}\,\sum_{\text{topologies}}\,\,\includegraphics[scale = .5,valign = c]{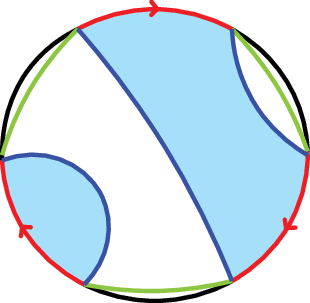},\\
	B_m&=\frac{\sqrt{\lambda}}{\sqrt{k_1 k_2}}\,\,\includegraphics[scale = .5,valign = c]{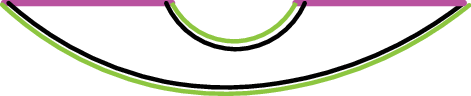} = \frac{\sqrt{\lambda}}{\sqrt{k_1 k_2}}\,\sum_{\substack{\text{connected} \\ \text{topologies}}}\,\,\includegraphics[scale = .5,valign = c]{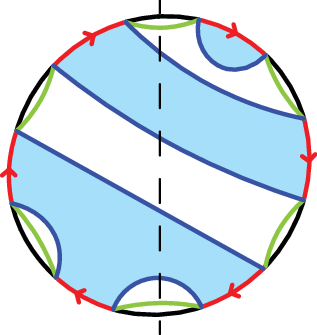},
\end{align}
where connected topologies are the ones which have geometries joining at least two asymptotic boundaries on either side of the vertical dashed line. We define the normalized quantities $\hat{B}_m=\frac{B_m}{Z_1}$ and $\hat{D}_m=\frac{D_m}{\sqrt{Z_1}}$. Further, these quantities are related by a sum rule: 
\begin{equation}
	\widehat{D}_m^2 + \widehat{B}_m = 1 
\end{equation}

The solution to the resolvent is then given by \cite{Akers:2021pvd}
\begin{equation}\label{eq:ref_spectrum} 
	R(\lambda) \approx \frac{(\chi_A\chi_B)^2}{2\widehat{B}_m\lambda}\left( \lambda + \widehat{B}_m\left( \frac{1}{\chi^2_B}-\frac{1}{\chi^2_A} \right) - \sqrt{(\lambda-\lambda_+)(\lambda-\lambda_-)} \right) + \frac{1}{\lambda-\widehat{D}^2_m}
\end{equation}
with 
\begin{equation}
	\lambda_\pm = \frac{\widehat{B}_m}{(k_1 k_2)^2}\left( \sqrt{k_1^2-1} \pm \sqrt{k_2^2-1} \right)^2 
\end{equation}

\begin{figure}
\centering 
		\includegraphics[width=0.5\linewidth]{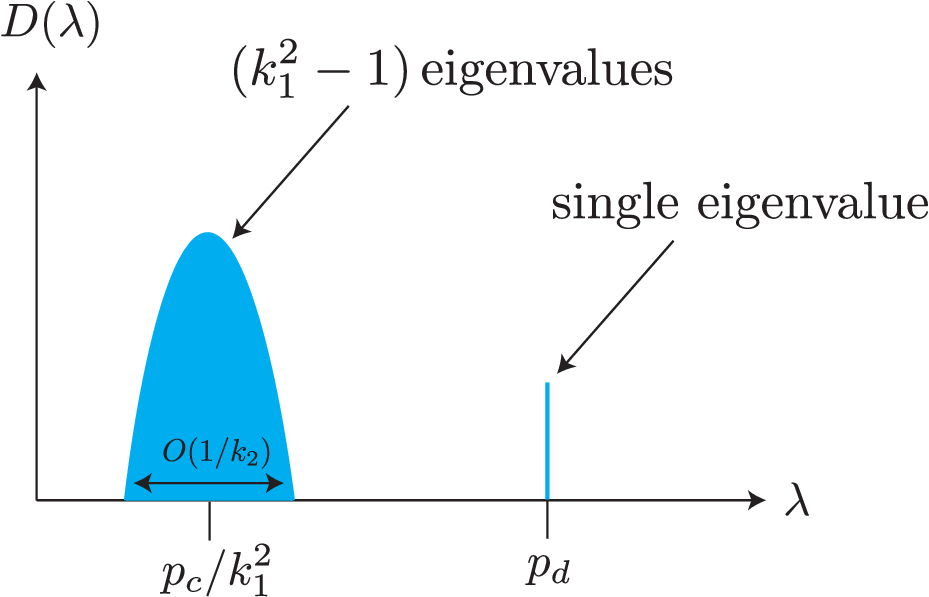} 
	\caption{The spectrum obtained from the resolvent \Eqref{eq:ref_spectrum}. We take $k_1<k_2$ here so the MP distribution has no zero eigenvalues.} 
\label{fig:spectrum} \end{figure}

The spectrum obtained from \Eqref{eq:ref_spectrum} takes a simple form (see \figref{fig:spectrum}), with two superselection sectors: a single pole of weight $p_d=\hat{D}_m^2$ and a mound of eigenvalues with weight $p_c = \hat{B}_m$. The mound is a Marchenko-Pastur (MP) distribution with support between $\lambda_{\pm} \simeq \widehat{B}_m(k_1^{-1}\pm k_2^{-1})^2$ and $\min(k_1^2-1,k_2^2-1)$ eigenvalues. I.e., the spectrum is given by 
\begin{align}
	D(\lambda) = \frac{(k_1 k_2)^2}{2\pi p_c\lambda}\sqrt{(\lambda_+-\lambda)(\lambda-\lambda_-)} + \delta(\lambda)(k_1^2-k_2^2)\theta(k_1^2-k_2^2) + \delta(\lambda-p_d),
\end{align}
where the sum rule then takes the form:
\begin{equation}
	p_c + p_d = 1,
\end{equation}
and thus we see that these quantities act like two ``classical probabilities''.

\begin{figure}[h!]
    \centering
    \includegraphics[width=0.8\textwidth]{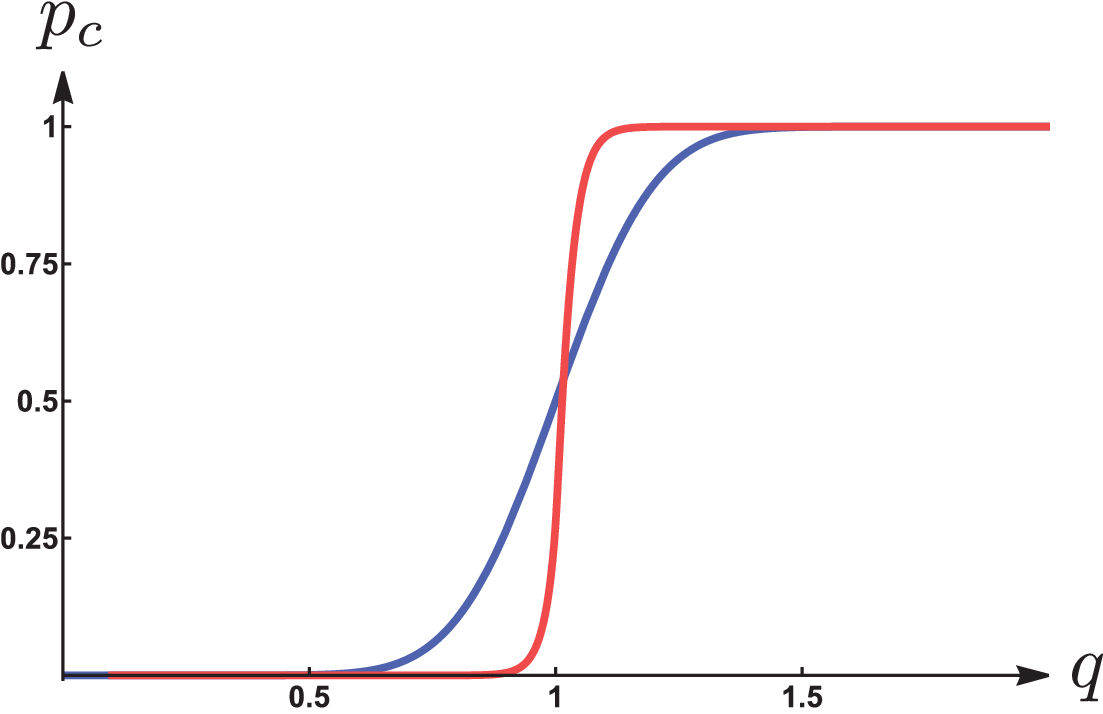}
    \caption{Plot of $p_c(q)\approx \frac{S_R(R_1:R_2)}{2 \min(S(R_1),S(R_2))}$ for $m=1$ as a function of $q=\frac{\beta s_k}{2\pi}$, for $\beta=1$, in the canonical ensemble (blue) and microcanonical ensemble (red). The phase transition in the canonical ensemble is spread out over a region of size $\Delta s = O(\frac{1}{\sqrt{\beta}})$ due to thermal fluctuations.}
    \label{fig:p01}
\end{figure}

\begin{figure}
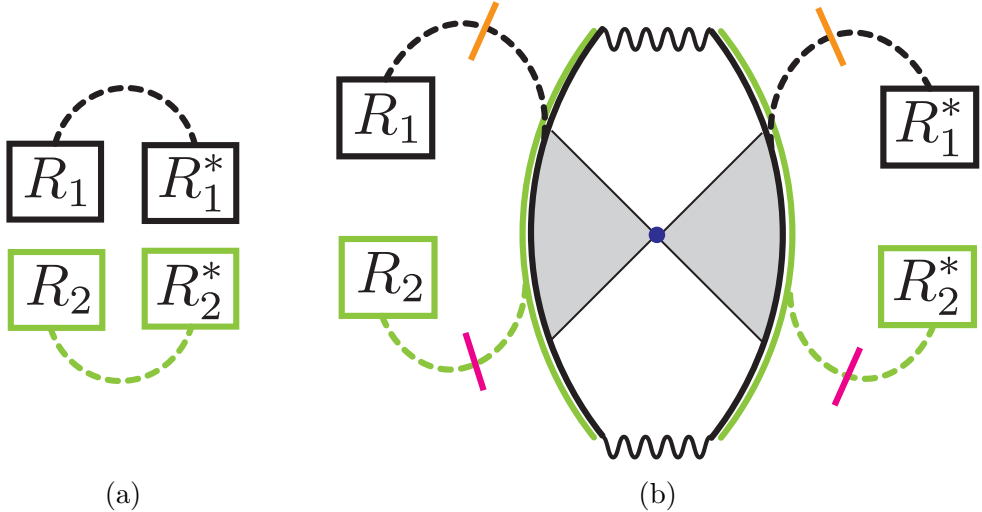

	\centering 
	\begin{subfigure}
		[b]{0.2
		\textwidth} 
		\raisebox{0.25\height}{\includegraphics[width=\textwidth]{fig2/disconnected_CP.eps}}
		\subcaption{} 
	\end{subfigure}
	\hspace{10mm} 
	\begin{subfigure}
		[b]{0.55
		\textwidth} 
		\raisebox{0pt}{\includegraphics[width=\textwidth]{fig2/connected_CP.eps}}
		\subcaption{} 
	\end{subfigure}
	\caption{(a) The canonical purification in the disconnected phase has radiation systems purifying themselves. (b) The canonical purification in the connected phase has a state of radiation systems entangled via a closed universe. The two candidate RT surfaces are denoted orange and pink.} 
\label{fig:CP_phases} \end{figure}

Using \Eqref{eq:bmdm}, we obtain
\begin{align}\label{eq:p0}
	p_d &= \frac{(\tr(\rho_{R_1 R_2}^{m/2}))^2}{k \tr{\rho_{R_1 R_2}^{m}}}\\
	p_c &= 1-p_d.
\end{align}
we interpret these superselection sectors as corresponding to the two different geometries shown in \figref{fig:CP_phases}, a disconnected and connected purification respectively. Thus, the reflected spectrum depends on the Renyi entropies of $\rho_{R_1 R_2}$, which we computed in \secref{sub:renyi}. For $m=1$, the physically most interesting case, we have from \Eqref{eq:p0} and \Eqref{eq:prob},
\begin{equation}
	p_d \approx k\lambda_0 \approx \int_{s_k}^{\infty}ds\,\rho(s)w(s),
\end{equation}
which can be interpreted as the Pre-Page probability as discussed earlier. Consequently, the connected probability $p_c$ is interpreted as the Post-Page probability. We plot this result and compare the  phase transitions in the canonical and microcanonical ensemble in \figref{fig:p01}. Due to thermal fluctuations, the phase transition is spread out over a range of parameter space of size $\Delta s = O(\frac{1}{\sqrt{\beta}})$. 

The reflected entropy is given by (assuming $k_1<k_2$) 
\begin{align}
	   	S_R(R_1:R_2) = \begin{cases}
		0 & k < \exp(S_{BH})\\
		2 \log k_1 &  k>\exp(S_{BH}).
	\end{cases}
\label{eq:main_result} \end{align}
This is the main result of the paper. The reflected entropy is given by:
\begin{align}
S_R(R_1:R_2)\approx p_c (2 \min (S(R_1),S(R_2))),     
\end{align}
where we generally have the upper bound $S_R(R_1:R_2)\leq 2 \min (S(R_1),S(R_2))$. Thus, \figref{fig:p01} serves as an approximate plot of $S_R(R_1:R_2)$ normalized by the upper bound.



\subsection{Renyi Reflected Entropy} 
\label{sub:renyi_sr}

Having discussed the reflected entropy, we now analyze the $(m,n)$-Renyi reflected entropies. Since the spectrum is an analytic function of $m$, the Renyi entropies can be computed by using the sum of the moments of the two sectors in the spectrum \cite{Akers:2021pvd}
\begin{equation}\label{eq:sn}
	S_n(R_1 R_1^*) = \frac{1}{1-n} \ln \left( p^n_d + k_1^{2(1-n)} r^{-2} C_n(r^2) p^n_c \right) 
\end{equation}
where $r=k_1^2/k_2^2$ and $C_n(q)$ is the $q$-Catalan number defined by
\begin{equation}
    C_n(q) = \sum^n_{k=0} q^k N(n,k),
\end{equation}
where $N(n,k)$ is the Narayana number. Notably, the probabilities are the only source of $m$-dependence. Thus, we need to analyze the behaviour of $p_d(m)$ as we cross the Page transition using the analysis in \secref{sub:renyi}.

\begin{figure}[h!]
    \centering
    \includegraphics[width=0.8\textwidth]{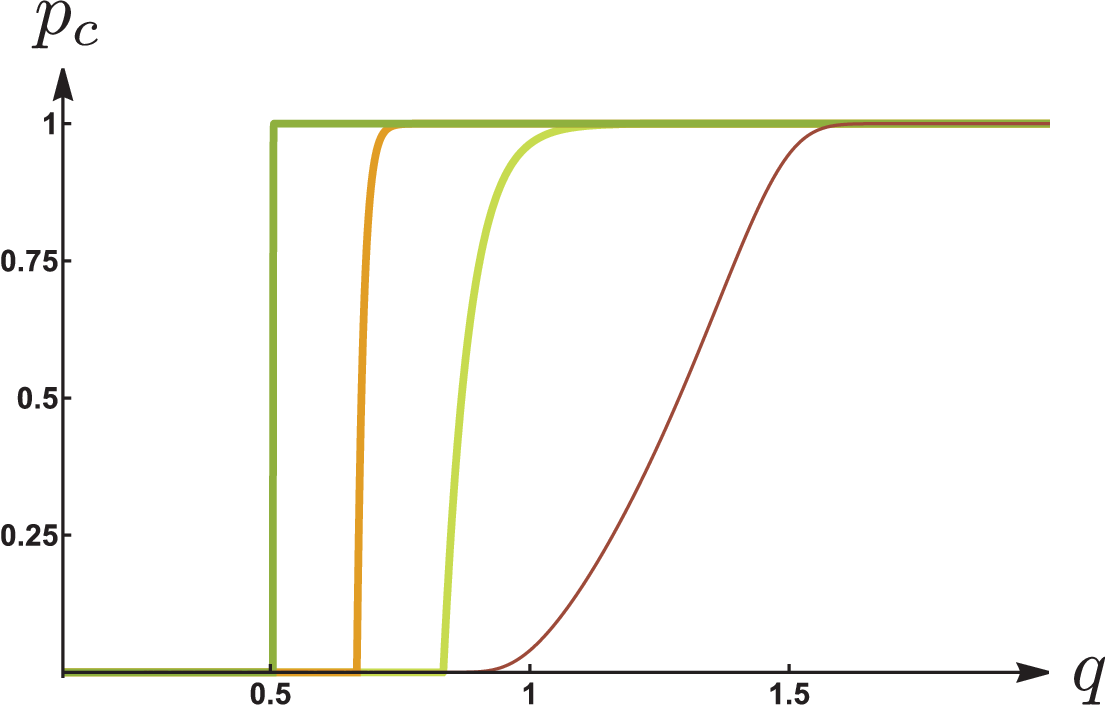}
    \caption{Analytic plot of $p_c(m,q)$ for $m=1.5$ (yellow), $m=3$ (orange) and $m=100$ (green) and numerical plot for $m=0.7$ (thin, brown) as a function of $q=\frac{\beta s_k}{2\pi}$ for $\beta=1$. The locations at which the $(m,1)$ Renyi reflected entropies undergo a transition changes as a function of $m$ for $m>1$.}
    \label{fig:p0m}
\end{figure}

For $m\geq2$, we obtain
\begin{equation}
	p_d \approx \begin{cases} 1  & s_k < s_k^{(m)}\\
		 \frac{\exp\left[(m-1)S_0+(m-\frac{1}{m})\frac{2\pi^2}{\beta}\right]}{k^{m-1}} & s_k^{(m)}< s_k < s_k^{(m/2)} \\
				\frac{\exp \left[S_0+\frac{6\pi^2}{m\beta}\right]}{k} & s_k^{(m/2)} < s_k.	\end{cases}
\end{equation}
There are two continuous transitions in $p_d$ and hence, also in the $(m,n)$-Renyi reflected entropy. However, the second transition is almost invisible since $p_d$ is already non-perturbatively small at $s_k=s_k^{(m/2)}$ and the reflected entropy is dominated by the connected entanglement wedge. 

For $m \in (1,2]$, the phase transition in $p_d$ is largely determined by the phase transition in $\tr(\rho_{R_1 R_2}^m)$ which occurs at $s_k=s_k^{(m)}<s_1$. Thus, we obtain
\begin{equation}
	p_d \approx \begin{cases} 1  & s_k < s_k^{(m)}\\
		 \frac{\exp\left[(m-1)S_0+(m-\frac{1}{m})\frac{2\pi^2}{\beta}\right]}{k^{m-1}} & s_k^{(m)}< s_k < s_1 \\
				O(e^{-\frac{1}{\beta}}) & s_1 < s_k,	\end{cases}
\end{equation}
where the value in the last line is difficult to compute analytically, although it is non-perturbatively suppressed and has a negligible effect on the reflected entropy. We plot the behavior for $p_d(m)$ for different values of $m>1$ in \figref{fig:p0m}.

Finally, for the case $m<1$, there is a multiplicative uncertainty in the calculation of $\tr(\rho_{R_1 R_2}^m)$ which makes it difficult to compute $p_d(q)$ analytically. We provide a numerical plot instead in \figref{fig:p0m}. Qualitatively, the transition happens over a large region in parameter space corresponding to $s_k\in[s_1,s_m]$.

To summarize, the phase transition in the $(m,n)$-Renyi reflected entropy comes from a shift in dominance between the two terms in \Eqref{eq:sn}. As $n\rightarrow 1$, the transition is dictated by the phase transition in $p_d$. This transition is qualitatively different for $m>1$ and $m<1$. For $m>1$, it is a sharp transition at $s_k=s_k^{(m)}$ and for $m<1$, it is spread out over $s_k\in[s_1,s_m]$.


\section{Summary and Discussion} 
\label{sec:disc}

In this paper we have analyzed the canonical purification and reflected entropy in the West Coast Model. We found that the holographic proposal is satisfied in the expected regimes: the canonical purification is related to a doubled spacetime, and the reflected entropy is related to the entanglement wedge cross section. Furthermore, by summing over the contributions from all relevant saddles, we have understood how the geometric picture evolves as we cross an entanglement phase transition. We now discuss certain interesting aspects of our calculation and potential directions for future research.

\subsection{Relation to Petz Map} 
\label{sub:petz}

Consider a code subspace of bulk states encoded in the boundary Hilbert space. The Petz map provides an explicit realization in this setting of a reconstruction that maps a given bulk operator to a boundary operator \cite{Cotler:2017erl,Chen:2019gbt,Penington:2019kki}. In the West Coast Model, one can consider bulk operators that live on the ETW brane. Before the Page transition, they are reconstructable by the black hole $B$, whereas after the Page transition they are reconstructable by the radiation system $R$. More so, in Ref.~\cite{Penington:2019kki}, they computed the probability of the Petz map acting on the radiation to succeed. It turns out to be given by $p_c$, precisely the probability that showed up in our calculation of the reflected spectrum (regardless of the choice of how to split $R$ into $R_1 R_2$). 
This is true in both the microcanonical ensemble and the canonical ensemble (see \figref{fig:p01} for our $p_c$, which matches the Petz recovery probability computed in Ref.~\cite{Penington:2019kki}).

This hints at a connection between the reflected entropy and the Petz map success probability.
From one point of view, this seems completely sensible within the paradigm of entanglement wedge reconstruction. The reflected entropy $S_R(R_1:R_2)$ is a sharp diagnostic of whether the entanglement wedge is connected, and only when it's connected does the entanglement wedge of $R_1 R_2$ include the black hole interior.
Indeed, the probability $p_c$ is simply the probability of measuring the canonical purification of $R_1 R_2$ in the connected phase. 
It will be interesting to investigate a possible deeper connection between the reflected entropy and reconstruction maps.


\subsection{Area Fluctuations} 
\label{sub:area_fluctuations}

\begin{figure}[h]
    \centering
    \includegraphics[width=0.4\textwidth]{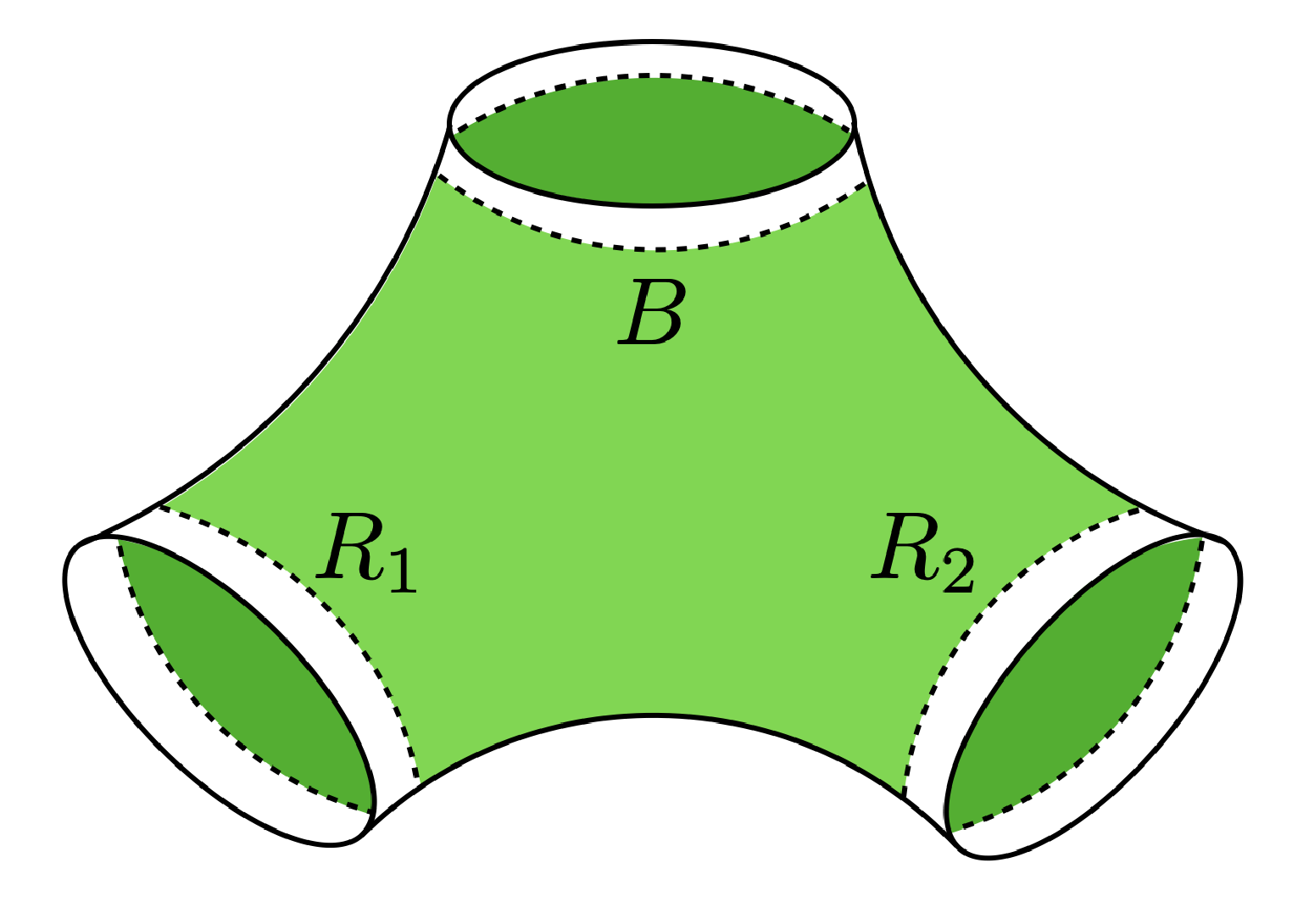}
    \caption{The three-boundary wormhole corresponding to our current setup.}
    \label{fig:3bdry}
\end{figure}

Here, we considered a holographic state in which the horizon area for $B$ had area fluctuations. However, in order to do the calculation, we needed to consider a maximally entangled state between $R$ and the ETW brane flavours. In the ER=EPR story, this can be interpreted as considering a three boundary wormhole with area fluctuations for the horizon of $B$ but fixed area boundary conditions for the horizons of $R_1$ and $R_2$ respectively, see Figure~\ref{fig:3bdry}. It would be interesting in the future to understand how the area fluctuations of $R_1$ and $R_2$ affect the reflected entropy. Presumably, it will affect the spectrum on $R_1 R_1^*$ by including the effects of thermal fluctuations into the corresponding superselection sectors. One may attempt to analyze this using the techniques of the equilibrium approximation \cite{Liu:2020jsv,Vardhan:2021mdy}.

\acknowledgments

We would like to thank Xi Dong, Jonah Kudler-Flam, Don Marolf, Sean McBride and Wayne Weng for useful discussions. PR is supported in part by a grant from the Simons Foundation, and by funds from UCSB. CA is supported by the Simons foundation as a member of the It from Qubit collaboration. This material is based upon work supported by the Air Force Office of Scientific Research under award number FA9550-19-1-0360.  TF is supported in part by Department of Energy under award number DE-SC0019183.

\bibliographystyle{jhep}
\bibliography{mybibliography}

\providecommand{\href}[2]{#2}\begingroup\raggedright\begin{thebibliography}{10}

\bibitem{Hawking:1974sw}
S.~W. Hawking, {\it {Particle Creation by Black Holes}},  {\em Commun. Math.
  Phys.} {\bf 43} (1975) 199--220. [,167(1975)].

\bibitem{Mathur:2009hf}
S.~D. Mathur, {\it {The Information paradox: A Pedagogical introduction}},
  {\em Class. Quant. Grav.} {\bf 26} (2009) 224001,
  [\href{http://arxiv.org/abs/0909.1038}{{\tt arXiv:0909.1038}}].

\bibitem{Almheiri:2012rt}
A.~Almheiri, D.~Marolf, J.~Polchinski, and J.~Sully, {\it {Black Holes:
  Complementarity or Firewalls?}},  {\em JHEP} {\bf 02} (2013) 062,
  [\href{http://arxiv.org/abs/1207.3123}{{\tt arXiv:1207.3123}}].

\bibitem{Almheiri:2013hfa}
A.~Almheiri, D.~Marolf, J.~Polchinski, D.~Stanford, and J.~Sully, {\it {An
  Apologia for Firewalls}},  {\em JHEP} {\bf 09} (2013) 018,
  [\href{http://arxiv.org/abs/1304.6483}{{\tt arXiv:1304.6483}}].

\bibitem{Penington:2019kki}
G.~Penington, S.~H. Shenker, D.~Stanford, and Z.~Yang, {\it {Replica wormholes
  and the black hole interior}},  \href{http://arxiv.org/abs/1911.11977}{{\tt
  arXiv:1911.11977}}.

\bibitem{Almheiri:2019qdq}
A.~Almheiri, T.~Hartman, J.~Maldacena, E.~Shaghoulian, and A.~Tajdini, {\it
  {Replica Wormholes and the Entropy of Hawking Radiation}},
  \href{http://arxiv.org/abs/1911.12333}{{\tt arXiv:1911.12333}}.

\bibitem{Penington:2019npb}
G.~Penington, {\it {Entanglement Wedge Reconstruction and the Information
  Paradox}},  \href{http://arxiv.org/abs/1905.08255}{{\tt arXiv:1905.08255}}.

\bibitem{Almheiri:2019psf}
A.~Almheiri, N.~Engelhardt, D.~Marolf, and H.~Maxfield, {\it {The entropy of
  bulk quantum fields and the entanglement wedge of an evaporating black
  hole}},  \href{http://arxiv.org/abs/1905.08762}{{\tt arXiv:1905.08762}}.

\bibitem{Engelhardt:2014gca}
N.~Engelhardt and A.~C. Wall, {\it {Quantum Extremal Surfaces: Holographic
  Entanglement Entropy beyond the Classical Regime}},  {\em JHEP} {\bf 01}
  (2015) 073, [\href{http://arxiv.org/abs/1408.3203}{{\tt arXiv:1408.3203}}].

\bibitem{Dong:2017xht}
X.~Dong and A.~Lewkowycz, {\it {Entropy, Extremality, Euclidean Variations, and
  the Equations of Motion}},  {\em JHEP} {\bf 01} (2018) 081,
  [\href{http://arxiv.org/abs/1705.08453}{{\tt arXiv:1705.08453}}].

\bibitem{Akers:2021fut}
C.~Akers and G.~Penington, {\it {Quantum minimal surfaces from quantum error
  correction}},  \href{http://arxiv.org/abs/2109.14618}{{\tt
  arXiv:2109.14618}}.

\bibitem{Dutta:2019gen}
S.~Dutta and T.~Faulkner, {\it {A canonical purification for the entanglement
  wedge cross-section}},  \href{http://arxiv.org/abs/1905.00577}{{\tt
  arXiv:1905.00577}}.

\bibitem{Akers:2019gcv}
C.~Akers and P.~Rath, {\it {Entanglement Wedge Cross Sections Require
  Tripartite Entanglement}},  \href{http://arxiv.org/abs/1911.07852}{{\tt
  arXiv:1911.07852}}.

\bibitem{Chandrasekaran:2020qtn}
V.~Chandrasekaran, M.~Miyaji, and P.~Rath, {\it {Including contributions from
  entanglement islands to the reflected entropy}},  {\em Phys. Rev. D} {\bf
  102} (2020), no.~8 086009, [\href{http://arxiv.org/abs/2006.10754}{{\tt
  arXiv:2006.10754}}].

\bibitem{Hayden:2021gno}
P.~Hayden, O.~Parrikar, and J.~Sorce, {\it {The Markov gap for geometric
  reflected entropy}},  \href{http://arxiv.org/abs/2107.00009}{{\tt
  arXiv:2107.00009}}.

\bibitem{Vidmar:2017pak}
L.~Vidmar and M.~Rigol, {\it {Entanglement Entropy of Eigenstates of Quantum
  Chaotic Hamiltonians}},  {\em Phys. Rev. Lett.} {\bf 119} (2017), no.~22
  220603, [\href{http://arxiv.org/abs/1708.08453}{{\tt arXiv:1708.08453}}].

\bibitem{Murthy:2019qvb}
C.~Murthy and M.~Srednicki, {\it {Structure of chaotic eigenstates and their
  entanglement entropy}},  {\em Phys. Rev. E} {\bf 100} (2019), no.~2 022131,
  [\href{http://arxiv.org/abs/1906.04295}{{\tt arXiv:1906.04295}}].

\bibitem{Dong:2020iod}
X.~Dong and H.~Wang, {\it {Enhanced corrections near holographic entanglement
  transitions: a chaotic case study}},  {\em JHEP} {\bf 11} (2020) 007,
  [\href{http://arxiv.org/abs/2006.10051}{{\tt arXiv:2006.10051}}].

\bibitem{Marolf:2020vsi}
D.~Marolf, S.~Wang, and Z.~Wang, {\it {Probing phase transitions of holographic
  entanglement entropy with fixed area states}},  {\em JHEP} {\bf 12} (2020)
  084, [\href{http://arxiv.org/abs/2006.10089}{{\tt arXiv:2006.10089}}].

\bibitem{Akers:2021pvd}
C.~Akers, T.~Faulkner, S.~Lin, and P.~Rath, {\it {Reflected entropy in random
  tensor networks}},  \href{http://arxiv.org/abs/2112.09122}{{\tt
  arXiv:2112.09122}}.

\bibitem{Maldacena:2001kr}
J.~M. Maldacena, {\it {Eternal black holes in anti-de Sitter}},  {\em JHEP}
  {\bf 04} (2003) 021, [\href{http://arxiv.org/abs/hep-th/0106112}{{\tt
  hep-th/0106112}}].

\bibitem{Engelhardt:2017aux}
N.~Engelhardt and A.~C. Wall, {\it {Decoding the Apparent Horizon:
  Coarse-Grained Holographic Entropy}},  {\em Phys. Rev. Lett.} {\bf 121}
  (2018), no.~21 211301, [\href{http://arxiv.org/abs/1706.02038}{{\tt
  arXiv:1706.02038}}].

\bibitem{Engelhardt:2018kcs}
N.~Engelhardt and A.~C. Wall, {\it {Coarse Graining Holographic Black Holes}},
  {\em JHEP} {\bf 05} (2019) 160, [\href{http://arxiv.org/abs/1806.01281}{{\tt
  arXiv:1806.01281}}].

\bibitem{Engelhardt:2022qts}
N.~Engelhardt and r.~Folkestad, {\it {Canonical Purification of Evaporating
  Black Holes}},  \href{http://arxiv.org/abs/2201.08395}{{\tt
  arXiv:2201.08395}}.

\bibitem{Akers:2020pmf}
C.~Akers and G.~Penington, {\it {Leading order corrections to the quantum
  extremal surface prescription}},  {\em JHEP} {\bf 04} (2021) 062,
  [\href{http://arxiv.org/abs/2008.03319}{{\tt arXiv:2008.03319}}].

\bibitem{Shapourian:2020mkc}
H.~Shapourian, S.~Liu, J.~Kudler-Flam, and A.~Vishwanath, {\it {Entanglement
  negativity spectrum of random mixed states: A diagrammatic approach}},
  \href{http://arxiv.org/abs/2011.01277}{{\tt arXiv:2011.01277}}.

\bibitem{Dong:2021oad}
X.~Dong, S.~McBride, and W.~W. Weng, {\it {Replica Wormholes and Holographic
  Entanglement Negativity}},  \href{http://arxiv.org/abs/2110.11947}{{\tt
  arXiv:2110.11947}}.

\bibitem{Akers:2018fow}
C.~Akers and P.~Rath, {\it {Holographic Renyi Entropy from Quantum Error
  Correction}},  {\em JHEP} {\bf 05} (2019) 052,
  [\href{http://arxiv.org/abs/1811.05171}{{\tt arXiv:1811.05171}}].

\bibitem{Dong:2018seb}
X.~Dong, D.~Harlow, and D.~Marolf, {\it {Flat entanglement spectra in
  fixed-area states of quantum gravity}},  {\em JHEP} {\bf 10} (2019) 240,
  [\href{http://arxiv.org/abs/1811.05382}{{\tt arXiv:1811.05382}}].

\bibitem{Dong:2019piw}
X.~Dong and D.~Marolf, {\it {One-loop universality of holographic codes}},
  {\em JHEP} {\bf 03} (2020) 191, [\href{http://arxiv.org/abs/1910.06329}{{\tt
  arXiv:1910.06329}}].

\bibitem{Lewkowycz:2013nqa}
A.~Lewkowycz and J.~Maldacena, {\it {Generalized gravitational entropy}},  {\em
  JHEP} {\bf 08} (2013) 090, [\href{http://arxiv.org/abs/1304.4926}{{\tt
  arXiv:1304.4926}}].

\bibitem{Dong:2016fnf}
X.~Dong, {\it {The Gravity Dual of Renyi Entropy}},  {\em Nature Commun.} {\bf
  7} (2016) 12472, [\href{http://arxiv.org/abs/1601.06788}{{\tt
  arXiv:1601.06788}}].

\bibitem{Cotler:2017erl}
J.~Cotler, P.~Hayden, G.~Penington, G.~Salton, B.~Swingle, and M.~Walter, {\it
  {Entanglement Wedge Reconstruction via Universal Recovery Channels}},  {\em
  Phys. Rev. X} {\bf 9} (2019), no.~3 031011,
  [\href{http://arxiv.org/abs/1704.05839}{{\tt arXiv:1704.05839}}].

\bibitem{Chen:2019gbt}
C.-F. Chen, G.~Penington, and G.~Salton, {\it {Entanglement Wedge
  Reconstruction using the Petz Map}},  {\em JHEP} {\bf 01} (2020) 168,
  [\href{http://arxiv.org/abs/1902.02844}{{\tt arXiv:1902.02844}}].

\bibitem{Liu:2020jsv}
H.~Liu and S.~Vardhan, {\it {Entanglement entropies of equilibrated pure states
  in quantum many-body systems and gravity}},  {\em P. R. X. Quantum.} {\bf 2}
  (2021) 010344, [\href{http://arxiv.org/abs/2008.01089}{{\tt
  arXiv:2008.01089}}].

\bibitem{Vardhan:2021mdy}
S.~Vardhan, J.~Kudler-Flam, H.~Shapourian, and H.~Liu, {\it {Mixed-state
  entanglement and information recovery in thermalized states and evaporating
  black holes}},  \href{http://arxiv.org/abs/2112.00020}{{\tt
  arXiv:2112.00020}}.

\end{thebibliography}\endgroup
 
\end{document}